\documentclass[12pt]{article}

\ifx\pdfoutput\undefined
\usepackage[dvips,bookmarks]{hyperref}
\else
\usepackage{hyperref}
\fi
\hypersetup{colorlinks=false,bookmarksopen,bookmarksnumbered,citecolor=blue,
   pdfstartview=FitH}

\usepackage{latexsym}
\usepackage{amssymb,amsfonts,amsmath}
\usepackage{graphicx} 
\usepackage{indentfirst}
\usepackage{bbm}

\parskip=\medskipamount

\newcommand {\cD}{{\cal D}}
\newcommand {\cE}{{\cal E}}
\newcommand {\cF}{{\cal F}}

\newcommand {\cH}{{\cal H}}

\newcommand {\cL}{{\cal L}}
\newcommand {\cM}{{\cal M}}
\newcommand {\cN}{{\cal N}}
\newcommand {\cO}{{\cal O}}

\newcommand {\cQ}{{\cal Q}}

\newcommand {\cT}{{\cal T}}
\newcommand {\cU}{{\cal U}}

\newcommand {\cW}{{\cal W}}

\def\a{\alpha}
\def\b{\beta}
\def\c{\chi}
\def\d{\delta}

\def\g{\gamma}

\def\k{\kappa}

\def\m{\mu}

\def\p{\pi}
\def\q{\theta}

\def\s{\sigma}

\def\D{\Delta}
\def\F{\Phi}
\def\J{\Psi}
\def\L{\Lambda}
\def\O{\Omega}

\def\S{\Sigma}
\def\U{\Upsilon}

\def\ri{{\rm i}}

\newcommand{\ad}{{\dot{\alpha}}}                           
\newcommand{\bd}{{\dot{\beta}}}                            
\newcommand{\ve}{\varepsilon}                            
\newcommand{\cDB}{{\bar\cD}}                            

\newcommand{\ab}{{\a\b}}

\newcommand{\pa}{\partial}                           
\newcommand{\hf}{\frac12}

\newcommand{\be}{\begin{equation}}
\newcommand{\ee}{\end{equation}}
\newcommand{\bea}{\begin{eqnarray}}
\newcommand{\eea}{\end{eqnarray}}
\newcommand{\non}{\nonumber}
\newcommand{\ba}{\begin{array}}
\newcommand{\ea}{\end{array}}

\newcommand{\1}{\underline{1}}
\newcommand{\2}{\underline{2}}

\newcommand{\bm}[1]{\mbox{\boldmath$#1$}}

\def\double #1{#1{\hbox{\kern-2pt $#1$}}}

\newcommand{\gd}{{\dot\g}}
\newcommand{\dd}{{\dot\d}}

\newcommand{\ts}{{\tilde{\s}}}

\newcommand{\CD}{{\cD}}

\newcommand{\teb}{{\bar{\theta}}}

\newcommand{\bsubeq}{\begin{subequations}}
\newcommand{\esubeq}{\end{subequations}}

\newcommand{\ul}{\underline}
\newcommand{\eps}{{\epsilon}}

\newcommand{\dalpha}{{\dot{\alpha}}}
\newcommand{\dbeta}{{\dot{\beta}}}

\newcommand{\N}{{\mathcal N}}
\newcommand{\eol}{\notag \\}
\newcommand{\rd}{\mathrm d}
\newcommand{\BcD}{{\bar \cD}}
\newcommand{\BCD}{{\bar \cD}}
\newcommand{\ID}{{\boldsymbol \partial}}

\begin{document}

\begin{titlepage}
\begin{flushright}
March 2011\\
\end{flushright}
\vspace{5mm}

\begin{center}
{\Large \bf
New higher-derivative couplings in 4D $\bm{\cN = 2}$ supergravity}
\\ 
\end{center}

\begin{center}

{\bf Daniel Butter and  Sergei M. Kuzenko }

\footnotesize{
{\it School of Physics M013, The University of Western Australia\\
35 Stirling Highway, Crawley W.A. 6009, Australia}}  ~\\
\texttt{dbutter,\,kuzenko@cyllene.uwa.edu.au}\\
\vspace{2mm}

\end{center}
\vspace{5mm}

\begin{abstract}
\baselineskip=14pt
Using the off-shell formulation for general $\cN=2$ supergravity-matter systems
developed in arXiv:0905.0063, we propose a construction to generate a restricted chiral superfield 
from any real weight-zero projective multiplet $\cL$. One can choose $\cL$ to be  composed 
of tensor multiplets, $\cO(2n)$ multiplets, with $n=2,3,\dots$, and polar hypermultiplets.
In conjunction with the standard procedure 
to induce a $\cN=2$ linear multiplet from any chiral weight-one scalar, we obtain a powerful
mechanism to generate higher-derivative couplings in supergravity. In the case that $\cL$ 
is a homogeneous function of $n$ tensor multiplets of degree zero, we show that our construction is 
equivalent to that developed by de Wit and Saueressig in arXiv:hep-th/0606148 using the component 
superconformal tensor calculus. We also work out nontrivial examples with $\cL$ composed of $\cO(2n)$
and tensor multiplets.
\end{abstract}

\vfill
\end{titlepage}

\newpage
\renewcommand{\thefootnote}{\arabic{footnote}}
\setcounter{footnote}{0}

\tableofcontents


\numberwithin{equation}{section}


\section{Introduction}
\setcounter{footnote}{0}

Recently, an off-shell formulation for general $\cN=2$ supergravity-matter couplings
in four space-time dimensions has been constructed  \cite{KLRT-M1,K-08,KLRT-M2}.
It is a curved-space extension of the superconformal projective multiplets and their couplings  \cite{K-hyper}
in 4D $\cN=2$ projective superspace \cite{KLR,LR1,LR2}.
In the present paper, we will demonstrate that 
the methods developed in \cite{KLRT-M1,K-08,KLRT-M2} and closely related works
\cite{KT-M,KT-M2} allow us to generate new off-shell higher-derivative couplings
in $\cN=2$ supergravity, a problem which has been of some interest recently 
\cite{AHNT,dWKvZ}.

The construction we pursue in this paper is based on a duality between two basic off-shell representations
of $\cN=2$  supersymmetry -- the vector multiplet \cite{GSW} and the tensor multiplet \cite{Wess}.
The vector multiplet can be described in curved superspace by its
covariantly chiral field strength $W$  subject to the Bianchi identity\footnote{Such a superfield 
is often called reduced chiral.} \cite{GSW,Howe}
\bea
\cDB^\ad_i W= 0~, \qquad
\S^{ij} := \frac{1}{ 4} \Big(\cD^{\a(i}\cD_\a^{j)}+4S^{ij}\Big) W&=&\frac{1}{ 4} 
\Big(\cDB_\ad{}^{(i}\cDB^{j) \ad}+4\bar{S}^{ij}\Big)\bar{W} ~,
\label{1.1}
\eea
where $S^{ij} $ and ${\bar S}^{ij} $  are special dimension-1 components of the torsion.
The superfield $\Sigma^{ij}$ is real, ${\bar \S}_{ij} :=(\S^{ij})^* =\ve_{ik}\ve_{jl}\S^{kl}$, and obeys 
the constraints
\begin{align}
\cD^{(i}_\a \Sigma^{jk)} =  {\bar \cD}^{(i}_\ad \Sigma^{jk)} = 0~.
\end{align}
These constraints are characteristic  of the $\cN=2$ linear multiplet \cite{BS,SSW}.\footnote{Our
 curved-superspace conventions follow  Ref. \cite{KLRT-M2}. In particular,  
we use Howe's superspace realization \cite{Howe}  (see also \cite{Muller})
of $\cN=2$ conformal supergravity in which 
the  structure group is ${\rm SL}(2,{\mathbb C})\times {\rm U}(2)$. 
The relevant information about Howe's formulation is collected in Appendix \ref{Appendix A}.
In what follows, we will use the notation:
$\cD^{ij}:= \cD^{\a(i}\cD_\a^{j)}$ and ${\bar \cD}^{ij} := \cDB_\ad{}^{(i}\cDB^{j) \ad}$.
It should be noted that Howe's realization of $\cN=2$ conformal supergravity 
\cite{Howe} is a simple extension of Grimm's formulation \cite{Grimm} with the structure group  
${\rm SL}(2,{\mathbb C})\times {\rm SU}(2)$. The precise relationship between these two formulations 
is spelled out  in \cite{KLRT-M2}.}

There are several ways to realize $W$ as a gauge invariant field strength.
One possibility is to introduce a curved-superspace extension 
of Mezincescu's prepotential \cite{Mezincescu} (see also \cite{HST}),  $V_{ij}=V_{ji}$,
which is an unconstrained real SU(2) triplet. The expression for $W$ in terms of $V_{ij}$ is
\begin{align}
W = \frac{1}{4}\bar\Delta \Big({\cD}^{ij} + 4 S^{ij}\Big) V_{ij}~,
\end{align}
where $\bar\Delta$ is the chiral projection operator \eqref{chiral-pr}. 
Within the  projective-superspace approach of \cite{KLRT-M1,K-08,KLRT-M2}, the constraints on
$W$ can be solved  in terms of a real weight-zero tropical prepotential $V(v^i)$. 
The solution  \cite{KT-M} is
\bea
&&W  = \frac{1}{8\pi}  \oint v^i {\rm d}v_i
 \Big( ({\bar \cD}^-)^2 +4 \bar{S}^{--}\Big) V(v)~, 
  \label{1.4}\\
 && \qquad \qquad  \CD_\alpha^+ V = \BCD_\dalpha^+ V = 0 ~, \qquad  V(c\,v^i)=V(v^i)~, \qquad
\breve{V} =V~. \non
\eea
Here $v^i \in {\mathbb C}^2 \setminus \{0\}$ denote
homogeneous coordinates for ${\mathbb C}P^1$, and
the contour integral is carried out around the origin in ${\mathbb C}P^1$; see Appendix C 
for our isotwistor notation and conventions, 
including  the definition of the covariant derivatives $\cD^\pm_\a$ and ${\bar \cD}^\pm_\ad$. 
We discuss the relations between these two formulations in Appendix \ref{Wprepotential}.

In the rigid supersymmetric case, the representation (\ref{1.4}) can be derived 
from a more general result in harmonic superspace \cite{Zupnik,GIOS}
\bea
W = \frac{1}{4} \int {\rm d}u \,({\bar D}^-)^2 V^{++}~,
\label{1.5-harmonic}
\eea
which is given in terms of  an analytic prepotential  $V^{++}  (u^+_i, u^-_j) $ for the vector multiplet \cite{GIKOS}.
Such a derivation makes use of 
the singular reduction procedure introduced in \cite{K-double}. 
This has been carried out explicitly in five space-time dimensions \cite{KL}.
Unfortunately, a curved-superspace generalization of (\ref{1.5-harmonic})
has not yet been found.

The tensor  (or {\it linear}) multiplet can be described in curved superspace by
its gauge invariant field strength $G^{ij}$  which is defined to be a  real ${\rm SU}(2)$ triplet (that is, 
$G^{ij}=G^{ji}$ and ${\bar G}_{ij}:=(G^{ij})^* = G_{ij}$)
subject to the covariant constraints  \cite{BS,SSW}
\bea
\cD^{(i}_\a G^{jk)} =  {\bar \cD}^{(i}_\ad G^{jk)} = 0~.
\label{1.2}
\eea
These constraints are solved in terms of a chiral
prepotential $\Psi$ \cite{HST,GS82,Siegel83,Muller86} via
\begin{align}
\label{eq_Gprepotential}
G^{ij} = \frac{1}{4}\Big( \cD^{ij} +4{S}^{ij}\Big) \Psi
+\frac{1}{4}\Big( \cDB^{ij} +4\bar{S}^{ij}\Big){\bar \Psi}~, \qquad
{\bar \cD}^i_\ad \J=0~,
\end{align}
which is invariant under shifts
$\Psi \rightarrow \Psi + \ri \Lambda$,
with $\Lambda$ a reduced chiral superfield.
Associated with $G^{ij}$ is the real $\cO(2)$ projective multiplet
$G^{++}(v):= G^{ij} v_iv_j$ \cite{KLRT-M1,K-08,KLRT-M2}.
The constraints (\ref{1.2}) are equivalent to
\bea
\cD^+_\a G^{++} = {\bar \cD}^+_\ad G^{++} =0~ , \qquad 
\cD^+_\a := v_i \cD^i_\a~, \quad {\bar \cD}^+_\ad := v_i {\bar \cD}^i_\ad~.
\eea

The above properties of vector and tensor multiplets are related and complementary. 
In fact, they  can be used to generate linear multiplets from reduced chiral ones  
and vice versa.
Consider a system of $n_V$ Abelian vector multiplets
described by covariantly chiral field strengths $W_I$, $I=1,\dots, n_V$. 
Let $F(W_I)$ be a holomorphic homogeneous function of degree one, $F(c\,W_I)= c\,F(W_I)$.
Then, we can define a composite linear multiplet
\be
{\mathbb G}^{ij}:=\frac{1}{4}\Big( \cD^{ij}+4S^{ij}\Big)F(W_I)
+ \frac{1}{4}\Big(  {\bar \cD}^{ij}+4{\bar S}^{ij}\Big){\bar F}({\bar W}_I)~.
\label{Sigma2}
\ee
Using the algebra of the covariant derivatives, one can check that $ {\mathbb G}^{ij}$
indeed obeys the constraints (\ref{1.2}) and that the construction is invariant
under shifts of the form
\begin{align}
F(W_I) \rightarrow F(W_I) + \ri \,r_J W_J~,
\end{align}
for real constants $r_J$. Eq. (\ref{Sigma2}) is a standard construction to generate composite linear multiplets.
Of course, the  construction (\ref{Sigma2}) is a trivial application of 
(\ref{eq_Gprepotential}).

Conversely, we may take a system of $n_T$ tensor multiplets described by their field strengths $G^{ij}_A$,  
with $A =1, \dots, n_T$,  and let 
$G^{++}_A:= v_i v_j G^{ij}_A $ be the corresponding
covariant $\cO(2)$ multiplets. Let $L(G^{++}_A)$ be a real homogeneous function 
of degree zero, $L(c\,G^{++}_A)= L(G^{++}_A)$, and thus $L(G^{++}_A)$ is a covariant real weight-zero 
projective multiplet. 
Then, we can generate a composite {\it reduced chiral} multiplet defined by
\bea
{\mathbb W}  &=& \frac{1}{8\pi}  \oint_C v^i {\rm d}v_i
 \Big( ({\bar \cD}^-)^2 +4 \bar{S}^{--}\Big) L (G^{++}_A)~.  
\label{1.5}
\eea
The integration in (\ref{1.5}) is carried over a closed contour $C$ in ${\mathbb C}^2 \setminus \{0\}$. 
Similarly to eq. (\ref{1.4}), the right-hand side of (\ref{1.5}) involves a constant isospinor $u_i$ chosen to obey 
the constraint $(v,u):= v^i u_i \neq 0$, but otherwise is completely arbitrary. 
Using the constraints 
\bea
\cD^+_\a L( G^{++} )= {\bar \cD}^+_\ad L( G^{++} )=0~, 
\eea
one can check that the right-hand side of (\ref{1.5}) is invariant 
under arbitrary {\it projective transformations} of the form:
\be
(u^i\,,\,v^i)~\to~(u^i\,,\, v^i )\,R~,~~~~~~R\,=\,
\left(\begin{array}{cc}a~&0\\ b~&c~\end{array}\right)\,\in\,{\rm GL(2,\mathbb{C})}~,
\label{projectiveGaugeVar}
\ee
and the same property holds for the right-hand side of (\ref{1.4}).
This invariance guarantees that (\ref{1.4}) and (\ref{1.5}) are independent of the isospinor $u_i$.
Of course, the reduced chiral construction (\ref{1.5}) is a simple application of (\ref{1.4}).

The relation (\ref{1.5}) is an example of applying our new construction 
to generate composite reduced chiral multiplets. 
This construction will be derived in the main body of the paper.
A rigid supersymmetric version of (\ref{1.5}) was described for a special case 
by Siegel twenty-five years ago \cite{Siegel85}.
More recently, it has been rediscovered and fully elaborated by de Wit and
Saueressig \cite{deWS} in a component approach using the superconformal tensor
calculus to couple it to conformal supergravity, but without a contour-integral representation.

This paper is organized as follows. In section 2 our main construction, eq. (\ref{2.11a}),
is derived. As a simple application, in section 3 we consider the improved tensor multiplet. 
We also analyse implications of fixing the super-Weyl gauge freedom with a tensor multiplet.
Section 4 is devoted to a general system of self-interacting tensor multiplet. 
We demonstrate that our superfield construction makes it possible 
to re-derive, in a simple way,  the key results of \cite{deWS}
obtained originally within the superconformal tensor calculus.
In section 5 we show how to construct reduced chiral multiplets
out of $\cO(2n)$ multiplets. 
In section 6 we discuss the new higher-derivative couplings and some other implications 
of our approach. The main body of the paper is accompanied by five technical appendices. 
In appendix A we give a summary of the superspace geometry for $\N=2$
conformal supergravity   introduced  originally in \cite{Howe} and elaborated 
in \cite{KLRT-M2}. Appendix B describes the properties of the $\cN=2$ chiral projection operator.
Appendix C contains our isotwistor notation and conventions for projective superspace.
In Appendix D we describe manifestly SU(2)-covariant techniques to evaluate contour integrals in
${\mathbb C}P^1$. Finally, Appendix E is devoted to two prepotential formulations for the $\cN=2$ 
vector multiplet.

\section{Main construction}
\setcounter{footnote}{0}
Within the formulation for $\cN=2$ supergravity-matter systems developed in  \cite{KLRT-M2}, 
the  matter fields are described in terms of covariant projective  multiplets. 
In addition to the local $\cN=2$ superspace
coordinates $z^{{M}}=(x^{m},\q^{\mu}_i,{\bar \q}_{\dot{\mu}}^i)$,
such a supermultiplet, $Q^{(n)}(z,v)$, depends on auxiliary  
isotwistor  variables $v^i \in  {\mathbb C}^2 \setminus  \{0\}$, 
with respect to which $Q^{(n)}$ is holomorphic and homogeneous,
$Q^{(n)}(c \,v) =c^n \,Q^{(n)}(v)$, on an open domain of ${\mathbb C}^2 \setminus  \{0\}$.
The integer parameter $n$ is called the weight of   $Q^{(n)}$.
In other words, such superfields are intrinsically defined in  ${\mathbb C}P^1$.
The covariant projective  supermultiplets are required to be 
annihilated by half of the supercharges, 
\be
\cD^+_{\a} Q^{(n)}  = {\bar \cD}^+_{\ad} Q^{(n)}  =0~, \qquad \quad 
\cD^+_{ \a}:=v_i\,\cD^i_{ \a} ~, \qquad
{\bar \cD}^+_{\dot  \a}:=v_i\,{\bar \cD}^i_{\dot \a} ~,
\label{ana-introduction}
\ee  
with $\cD_{{A}} =(\cD_{{a}}, \cD_{{\a}}^i,\cDB^\ad_i)$ the covariant superspace 
derivatives. The dynamics of supergravity-matter 
systems are described by locally supersymmetric actions of the form \cite{KLRT-M2}:
\bea
S (\cL^{++}) &=&
\frac{1}{2\pi} \oint_C v^i \rd v_i
\int \rd^4 x \,{\rm d}^4\q \,{\rm d}^4{\bar \q}
\,E\, \frac{{ W}{\bar { W}}\cL^{++}}{({ \S}^{++})^2}~, 
\qquad E^{-1}= {\rm Ber}(E_A{}^M)~,~~~
\label{InvarAc}
\eea
with ${ \S}^{++}(v):=\S^{ij}v_i v_j$.
Here the Lagrangian $\cL^{++}(z,v)$ is a covariant real projective 
multiplet of weight two.
The vector multiplet $W$ is used as
a supergravity compensator and $\Sigma^{ij}$ is defined as in \eqref{1.1}.
As shown in \cite{KT-M2}, 
the action \eqref{InvarAc} can be written as an integral over the chiral subspace
\bea
S(\cL^{++}) &=& \int {\rm d}^4x \,{\rm d}^4 \q \, \cE \, W \,{\mathbb W}~, \non \\
{\mathbb W}  &=& \frac{1}{8\pi}  \oint_C v^i \rd v_i\,
 \Big( ({\bar \cD}^-)^2 +4 \bar{S}^{--}\Big) {\mathbb V}~, 
 \qquad {\mathbb V}:=  \frac{\cL^{++}}{\S^{++}}~.
\label{2.2}
\eea
It was also proved in \cite{KT-M2} that $\mathbb W$ is a restricted chiral superfield.

We now give an alternative chiral representation for  the action 
 \eqref{InvarAc}
in the case that $\cL^{++}$ has  the form 
\be
\cL^{++} = G^{++} \cL~,
\label{2.4bas}
\ee
where $G^{++}$  is a tensor multiplet, and $\cL$ is a real weight-zero projective multiplet,
\begin{align}
\cD^+_\a \cL = {\bar \cD}^+_\ad \cL =0 ~, 
\qquad \cL(c\, v^i) = \cL(v^i)~, \qquad \breve{\cL}=\cL~. 
\end{align}
Rewriting $G^{++}$ in terms of its chiral potential $\Psi$ \eqref{eq_Gprepotential},
\bea
G^{++} (v)= \frac{1}{4}\Big( (\cD^{+})^2+4{S}^{++}\Big) \J
+\frac{1}{4}\Big( (\cDB^{+})^2+4\bar{S}^{++}\Big){\bar \J}~, \qquad
{\bar \cD}^i_\ad \J=0~.
\label{G++}
\eea
one may rearrange \eqref{InvarAc} into the form
\bea
S  (G^{++} \cL ) &=&
\frac{1}{2\pi} \oint_C v^i \rd v_i
\int \rd^4 x \,{\rm d}^4\q {\rm d}^4{\bar \q}
\,E\, \frac{\J{\bar { W}}\cL }{ { \S}^{++} } ~+~{\rm c.c.}
\label{2.88}
\eea
This can be reduced to a chiral integral by acting with the chiral projector
$\bar\Delta$ defined by eq. (\ref{chiral-pr})
(see \cite{KT-M2} for a detailed derivation)
\bea
S  (G^{++} \cL ) &=&
\frac{1}{2\pi} 
\int \rd^4 x \,{\rm d}^4\q \,\cE \,{\bar \D}
\oint_C v^i \rd v_i
\, \frac{\J{\bar { W}}\cL }{ { \S}^{++} } ~+~{\rm c.c.}
\eea
Now, it only remains to make use of 
an important representation discovered in \cite{KT-M2}.
Specifically, given an arbitrary isotwistor superfield $U^{(-2)}(z,v)$
of weight $-2$
(see \cite{KLRT-M1} for the definition of isotwistor supermultiplets), 
it was  shown in \cite{KT-M2} that
\bea
\bar{\D} \oint v^i \rd v_i \,U^{(-2)}&=&
\frac{1}{16}  \oint v^i \rd v_i
\Big( ({\bar \cD}^-)^2 +4\bar{S}^{--}\Big) 
\Big( ({\bar \cD}^+)^2 +4\bar{S}^{++}\Big) U^{(-2)}~.~~~~~~
\label{chiralproj2}
\eea
Applying this representation to (\ref{2.88}) gives
\bea
S  (G^{++} \cL ) &=&
\int \rd^4 x \,{\rm d}^4\q \, \cE \,\J {\mathbb W} ~+~{\rm c.c.}~,
\label{2.10a} 
\eea
where we have introduced the following composite  chiral superfield: 
\bea
{\mathbb W}  &=& \frac{1}{8\pi}  \oint_C v^i {\rm d}v_i
 \Big( ({\bar \cD}^-)^2 +4 \bar{S}^{--}\Big) \cL   ~,  \qquad {\bar \cD}_i^\ad {\mathbb W}=0~.
 \label{2.11a}
\eea
Because $\Psi$ is defined only up to gauge transformations
\bea
\d \J = {\rm i}\,\L ~, \qquad
\cDB^\ad_i \L= 
\Big(\cD^{\a(i}\cD_\a^{j)}+4S^{ij}\Big) \L-
\Big(\cDB_\ad{}^{(i}\cDB^{j) \ad}+4\bar{S}^{ij}\Big)\bar{\L} =0~,~~~
\eea
the superfield $\mathbb W$ must indeed be reduced chiral,
\bea
\Big(\cD^{\a(i}\cD_\a^{j)}+4S^{ij}\Big) {\mathbb W} &=&
\Big(\cDB_\ad{}^{(i}\cDB^{j) \ad}+4\bar{S}^{ij}\Big)\bar{\mathbb W} ~.
\eea
We present a more direct proof of this result in Appendix \ref{Wprepotential}.

\section{The improved tensor multiplet}
\setcounter{footnote}{0}
The improved $\cN=2$ tensor multiplet 
 \cite{deWPV,LR} is a unique theory of the $\cN=2$ tensor multiplet 
 which is superconformal in the rigid supersymmetric case, and super-Weyl invariant 
 in the presence of conformal supergravity. This theory is a natural generalization 
 of the improved $\cN=1$ tensor multiplet \cite{deWR}.
 It occurs as a second compensator in the minimal formulation of $\cN=2$ Poincar\'e 
 supergravity proposed by de Wit, Philippe and Van Proeyen \cite{deWPV}.
 Within  the projective-superspace formulation for $\cN=2$ supergravity  \cite{KLRT-M1,K-08,KLRT-M2}, 
 the Lagrangian for the tensor compensator is 
\begin{align}
\cL^{++} = G^{++} \ln \big(G^{++} / \ri \Upsilon^{+} \breve\Upsilon^{+}\big) ~,
\end{align}
see \cite{K-08} for more details.
Here $\U^+$ is a weight-one arctic hypermultiplet, and $\breve{\U}{}^+$ its smile-conjugate. 
As shown in \cite{K-08,KT-M2},  the superfields $\U^+$ and $\breve{\U}{}^+$
are pure gauge degrees of freedom in the sense that they do not contribute to
the tensor compensator action. 
Using the formulation \eqref{2.10a} and (\ref{2.11a}),
the tensor compensator action can be 
rewritten as a chiral action of the form 
\begin{align}
S = \int \rd^4 x \,{\rm d}^4\q \, \cE \,\J {\mathbb W} ~+~{\rm c.c.} ~,
\end{align}
where $\mathbb W$ denotes the following reduced chiral scalar
\begin{align}
\mathbb W &= \frac{1}{8\pi} \oint_C v^{i} \rd v_i \, \Big((\BcD^-)^2 + 4 \bar S^{--}\Big)
     \ln \big(G^{++} / \ri \Upsilon^{+} \breve\Upsilon^{+}\big)     ~.
     \label{3.3}
\end{align}
It can be seen that  the arctic multiplet $\U^+$ and its conjugate $\breve{\U}{}^+$
do not contribute to the contour integral, and so they will be ignored below.
Our goal in this section is to evaluate (\ref{3.3}) as well as to study the properties of 
$\mathbb W$.

\subsection{Evaluation of the contour integral}
\label{Evaluation of the contour integral}
We begin by evaluating the derivatives in (\ref{3.3}):
\begin{align}
\mathbb W &= \frac{1}{8\pi} \oint_C v^{i} \rd v_i \, \left(
     \frac{(\BcD^-)^2 G^{++}}{G^{++}}
     - \frac{\BcD_\dalpha^- G^{++} \BcD^{\dalpha -} G^{++}}{(G^{++})^2}
     + 4 \bar S^{--} \ln G^{++} \right)~.
     \label{3.4}
\end{align}
It follows from first principles that (\ref{3.3}) is invariant under the projective transformations 
(\ref{projectiveGaugeVar}) and therefore independent of $u_i$.
We wish to show explicitly 
that all of the $u$-dependence
in the integrand (\ref{3.4}) can be eliminated using the properties of the tensor multiplet.
First, using the constraints (\ref{1.2})
and the (anti)commutation relations (\ref{a-c2}),
one can show that 
\begin{subequations}
\bea
\BcD_\dalpha^- G^{++} 
  = \frac{2}{3} \, \bar \chi_\dalpha^{+} & \equiv  & \frac{2}{3}  {\bar \c}_{\ad}^iv_i
     ~,     \\
(\BcD^-)^2 G^{++}
     + 8 \bar S^{--} G^{++}
     - 4 \bar S^{-+} \ID^{--} G^{++}
     &=& \frac{1}{3} \bar M~,
\eea 
\end{subequations}
where we have defined 
\begin{subequations}
\bea \label{eq_defchi}
{\bar \c}^{\ad i}&:=&
\BcD^\dalpha_k G^{k i} ~, \\ \label{eq_defM}
\bar M &:=& \left(\BcD_{jk} + 12 \bar S_{jk} \right) G^{jk}~.
\eea
\end{subequations}
Using these formulae, we rewrite $\mathbb W$ as
\begin{align}
\mathbb W
     &= \frac{1}{8\pi} \oint_C v^{i} \rd v_i \, \left(
     \frac{1}{3} \frac{\bar M}{G^{++}}
     - \frac{4}{9} \frac{\bar\chi^+ \bar\chi^+}{(G^{++})^2}
+4\ID^{--} \Big( \bar S^{-+} \log G^{++}\Big)- 8 \bar S^{--}
     \right)~.
\end{align}
The terms involving $\bar S$ explicitly in this expression can be
rearranged into
\begin{align}
\frac{1}{8\pi} \oint_C v^{i} \rd v_i \, \ID^{--}\Big(  
      4 \bar S^{-+} \log G^{++}     - 8 \bar S^{-+}      \Big)
\end{align}
and so they vanish using integration by parts (see Appendix \ref{residue}
for technical details). We are left with
\begin{align}
\mathbb W
     &= \frac{1}{8\pi} \oint_C v^{i} \rd v_i \, \left(
     \frac{1}{3} \frac{\bar M}{G^{++}}
     - \frac{4}{9} \frac{\bar\chi^+ \bar\chi^+}{(G^{++})^2}\right)~.
\label{3.9}
\end{align}

The contour integral in (\ref{3.9}) can be evaluated in a manifestly SU(2) covariant way using the 
technique described in Appendix \ref{residue}.
Making use of  the result \eqref{eq_contour2}  leads to
\begin{align}\label{eq_I1}
\mathbb W &=
     - \frac{1}{24 G} \bar M
     + \frac{1}{36 G^3} \bar \chi^i \bar \chi^j G_{ij} ~,
\end{align}
where
\begin{align}
G^2 := \frac{1}{2} G^{ij} G_{ij}~.
\end{align}
This expression (up to normalizations) was discovered originally in \cite{deWPV}
using the superconformal tensor calculus.
It was  later reconstructed  in curved superspace
 \cite{Muller86} with the aid of the results in  \cite{deWPV} and \cite{Siegel85}.
 Its contour origin was explored in the globally supersymmetric case  by Siegel \cite{Siegel85}. 
Here we have derived it from superfield supergravity
and shown its  contour integral origin, eq.  (\ref{3.3}),  for the first time.

A curious feature of the reduced chiral superfield  \eqref{eq_I1} is that it can be rewritten in
a more elegant and compact form
\begin{align}\label{eq_imptensor}
\mathbb W &=
     -\frac{G}{8} (\BcD_{ij} + 4 \bar S_{ij}) \left(\frac{G^{ij}}{G^2} \right) ~.
\end{align}
Its rigid supersymmetric version appeared in our recent 
work on $\cN=2$ supercurrents \cite{Butter:2010sc}.
It is a laborious task to check that this does indeed match \eqref{eq_I1} in
supergravity by straightforwardly applying the properties of the tensor multiplet.
We will demonstrate the  equivalence of (\ref{eq_imptensor}) to \eqref{eq_I1} 
via an indirect route, through the
use of super-Weyl transformations \cite{Howe,KLRT-M2}.
Using the representation (\ref{eq_imptensor}), it is easy to show 
that the super-Weyl transformation law of  $\mathbb W$ 
coincides with that of the vector multiplet field strength.

\subsection{Super-Weyl gauge fixing with a tensor multiplet}
\label{seubsection3-2}
Within Howe's formulation for $\cN=2$ conformal supergravity \cite{Howe}, 
the super-Weyl transformation law of the spinor covariant derivatives is 
\begin{subequations}
\bea
\d \cD{}_\a^i&=&\hf {\s}\cD_\a^i+2(\cD^{\g i}\s)M_{\g\a}-2(\cD_{\a k}\s)J^{ki}
-\hf (\cD_\a^i\s) \,{\mathbb J} ~,
\label{Finite_D}\\
\d \cDB_{\ad i}&=&\hf {\s} \cDB_{\ad i}+2(\cDB^{\gd}_{i}\s)\bar{M}_{\gd\ad}
+2(\cDB_{\ad}^{k}\s)J_{ki}
+\hf(\cDB_{\ad i}\s)\,{\mathbb J} ~,
\label{Finite_Db} 
\eea
\end{subequations}
where the transformation parameter $\s$ is real  unconstrained, and $M_{\ab}$, $J_{ij}$ and $\mathbb J$ 
denote the Lorentz, SU$(2)_R$ and U$(1)_R$ generators respectively, 
see \cite{KLRT-M2} and Appendix \ref{Appendix A} for more details.
The super-Weyl transformations of the torsion and the curvature are given in \cite{KLRT-M2}.
The field strengths of the vector multiplet  $W $ and the tensor multiplet  $G^{ij} $
transform as primary fields,  
\begin{subequations}
\bea
\d_\s W &=&   \s W~, \\
\d_\s G^{ij} &=& 2 \s G^{ij}~.
\eea
\end{subequations}
The composite reduced chiral superfield (\ref{eq_imptensor}) transforms 
as the vector multiplet field strength.

In $\cN=2$ supergravity, one of the two conformal compensators 
is usually a vector multiplet $W$ (see, e.g., \cite{deWPV} for more details), 
$W\neq 0$.
Usually, the super-Weyl invariance is fixed by imposing a condition on $W$.
Specifically, one may use  the super-Weyl and local U$(1)_R$  freedom to 
impose the gauge condition
\begin{align}\label{eq_Wcompensator}
W = 1~.
\end{align}
This amounts to switching off the spinor U$(1)_R$ connections, 
\bea
\F^i_\a =\F^\ad_i =0~,
\label{W=1-a}
\eea
along with certain restrictions \cite{Howe, KLRT-M2} on the torsion,
namely
\begin{align}
G_{\alpha \dbeta}^{ij} = 0 ~,\qquad
S^{ij} = \bar S^{ij} ~,
\label{W=1-ab}
\end{align}
as well as identifying the torsion $G_{\alpha \dbeta}$ with
the U$(1)_R$ vector connection
\begin{align}
G_{\alpha \dbeta} = \Phi_{\alpha \dbeta} ~.
\label{W=1-c}
\end{align}

If the second conformal compensator is a tensor multiplet $G^{ij}$, with $G^2 \neq 0$,
we may instead choose this multiplet to fix the super-Weyl gauge freedom.
Specifically, we may  use the super-Weyl gauge freedom to impose the condition
\begin{align}
G^2 = \frac{1}{2} G^{ij} G_{ij} = 1 ~.
\label{3.18}
\end{align}
This fixes $G^{ij}$ to be a unit isovector  subject to the tensor multiplet constraints \eqref{1.2}.
Actually, these constraints are sufficient to enforce that $G^{ij}$
is covariantly chiral and antichiral. To prove the former, note that
 \eqref{1.2} is equivalent to 
\begin{align}\label{eq_temp1}
\CD_\alpha^i G_{jk} = \frac{2}{3} \delta^i_{(j} \CD^m G_{k) m}~.
\end{align}
It follows that
\begin{align}\label{eq_temp2}
G_{ij} \CD_\alpha^j G^2 = G_{ij} G^{kl} \CD_\alpha^j G_{kl}
     = \frac{2}{3} G^2 \, \CD_\alpha^j G_{ji}~.
\end{align}
When we set $G^2=1$, \eqref{eq_temp2} must vanish,
which in turn implies that \eqref{eq_temp1} must vanish as well.
Because $G^{ij}$ is real, we may conclude that it must be both
chiral \emph{and} antichiral in this gauge.

These conditions have a number of interesting consequences for the
superspace geometry. 
In accordance with (\ref{a-c1}),  the consistency condition for $G^{ij}$ being
chiral reads
\begin{align}
0 = \{\CD_\alpha^i, \CD_\beta^j\} G^{kl} &= 
     2 \eps_{\alpha \beta} \eps^{ij} S^{mn} J_{mn} G^{kl} + 4 Y_{\alpha \beta} J^{ij} G^{kl}~.
\end{align}
This condition and its conjugate  imply
\begin{gather}
Y_{\alpha \beta} = 0 ~,\qquad
\bar Y_{\dalpha \dbeta} = 0 ~;\\
S^{ij} \propto G^{ij} ~,\qquad
\bar S^{ij} \propto G^{ij}~.
\end{gather}
Because $Y_{\alpha \beta}$ vanishes, $S^{ij}$ is now completely antichiral
\begin{align}
\CD_\alpha^k S^{ij} = 0 ~,
\end{align}
due to the dimension-3/2 Bianchi identities \eqref{eq_dim3.5bianchi}.
We conclude that
\begin{align}
S^{ij} = -\bar \phi \,G^{ij} ~,\qquad
\bar S^{ij} = -\phi \,G^{ij}~,
\label{3.25}
\end{align}
where $\phi$ is some chiral scalar. 

Furthermore, in accordance with eq. (\ref{a-c2}),
the consistency condition for $G^{ij}$ being \emph{both}
chiral and antichiral reads
\begin{align}
0 = \{\CD_\alpha^i, \BCD_{\dbeta j} \} G^{kl}
     = -2i \delta^i_j \CD_{\alpha\dbeta} G^{kl} 
     + 8 G_{\alpha \dbeta} J^i_j G^{kl}
     - 4i \delta^i_j G_{\alpha \dbeta}^{mn} J_{mn} G^{kl}~.
\end{align}
This implies
\begin{align}
G_{\alpha \dalpha} = 0~,
\end{align}
as well as
\begin{align}
\CD_{\alpha \dalpha} G^{ij} = 4 G_{\alpha \dalpha}^{k(i} G_k^{j)}~.
\end{align}
This last condition is quite interesting. If we now use the ${\rm SU}(2)_R$ gauge
freedom to fix $G^{ij}$ to a \emph{constant} unit isovector, we find
\begin{align}
-2 \Phi_{\alpha \dalpha}{}^{k (i} G^{j)}_k = 4 G_{\alpha \dalpha}^{k(i} G_k^{j)}
\end{align}
which implies that the vector ${\rm SU}(2)_R$ connection is
$-2 G_{\alpha \dalpha}^{ij}$ up to terms which are proportional to
$G^{ij}$.

Thus we have some rather interesting structure.\footnote{An analogue of the 
super-Weyl gauge (\ref{3.18}) naturally occurs in $\cN=3$ and $\cN=4$
supergravity in three dimensions \cite{KLT-M}. Implications of such a gauge fixing are highly 
nontrivial in the case of $\cN=4$ supergravity.} Imposing the super-Weyl gauge  (\ref{3.18})
eliminates $G_{\alpha \dalpha}$ and $Y_{\alpha \beta}$,  as well as 
implying the relation (\ref{3.25})
for chiral $\phi$. In fact, the field $\phi$ is actually {\it reduced chiral}.
Making use of the algebra
of covariant derivatives as well as the dimension-3/2 Bianchi
identities \eqref{eq_dim3.5bianchi}, one may show that (in this gauge)
\begin{align}
\CD_{ij} \bar S^{kl} - \BCD_{ij} S^{kl} &= 
     - 4 S_{ij} \bar S^{kl} + 4 S^{kl} \bar S_{ij}  ~.
\end{align}
Contracting with $G_{kl} / 2$ gives
\begin{align}
\CD_{ij} \phi - \BCD_{ij} \bar \phi &= 
     - 4 S_{ij} \phi + 4 \bar S_{ij} \bar \phi 
\end{align}
which shows $\phi$ to indeed be reduced chiral.

This formulation allows us to show that \eqref{eq_imptensor}
is indeed \eqref{eq_I1} and that this is reduced chiral.
Equality follows since both expressions have identical super-Weyl
transformation properties and reduce in the gauge $G^2 = 1$ to
\begin{align}
\mathbb W = - \frac{1}{2} \bar S_{ij} G^{ij} = \phi ~.
\end{align}
Moreover, since $\phi$ is reduced chiral, both
\eqref{eq_I1} and \eqref{eq_imptensor} must be as well, since 
$\mathbb W$ changes as a primary field under the super-Weyl transformations, 
\bea
\d_\s {\mathbb W} &=& \s {\mathbb W}~.
\eea

\section{Self-interacting tensor multiplets}
\label{several_tensor_multiplets}
The improved tensor multiplet  is the unique super-Weyl invariant  action
available 
which involves a single tensor multiplet. In the case of
several tensor multiplets $G_A^{++}$, we can consider a locally supersymmetric and super-Weyl 
invariant  action generated by the Lagrangian
\begin{align}\label{eq_genTaction}
\cL^{++} = G_A^{++} \cF^A(G_B^{++})~,
\end{align}
where $\cF^A$ is a homogeneous function of degree zero,
$\cF^A(c \,G_B^{++}) = \cF^A(G_B^{++})$. 
This Lagrangian is simply a superposition of several terms of the form (\ref{2.4bas}).
Associated with this
Lagrangian are reduced chiral superfields
\begin{align}\label{eq_Wcontour}
\mathbb W^A = \frac{1}{8\pi}  \oint_C v^i {\rm d} v_i
 \Big( ({\bar \cD}^-)^2 +4 \bar{S}^{--}\Big) \cF^A(G^{++}_B)~.
\end{align}
Evaluating the derivatives on $\cF^A$ is a nearly identical
procedure to what we considered in the previous section with a single
tensor superfield. It is straightforward to find
\begin{align}\label{eq_Wcomponent}
\mathbb W^A = \frac{1}{3} \cF^{A, B} \bar M_B
     + \frac{4}{9} \cF^{A, B, C}{}_{ij} \, \bar\chi^i_B \bar \chi^j_C~,
\end{align}
where $\bar \chi^i_B$ and $\bar M_B$ are defined as in \eqref{eq_defchi}
and \eqref{eq_defM}, with $G^{ij} $ replaced by $G_B^{ij}$, and
\begin{align}\label{eq_F1}
\cF^{A, B} &\equiv \frac{1}{8\pi}  \oint_C v^i {\rm d} v_i \, \frac{\partial \cF^A}{\partial G^{++}_B}~, \\
\label{eq_F2a}
\cF^{A, B, C}{}_{ij} &\equiv
     \frac{1}{8\pi}  \oint_C v^k {\rm d} v_k \,v_i v_j
     \frac{\partial^2 \cF^A}{\partial G^{++}_B \partial G^{++}_C}~.
\end{align}
It is worth noting that the second of these expressions can be written as
\begin{align}\label{eq_F2b}
\cF^{A, B, C}{}_{ij} = \frac{\partial \cF^{A, B}}{\partial G_C^{ij}}~.
\end{align}
The reduced chiral superfield \eqref{eq_Wcomponent} was first constructed
a few years ago by de Wit and Saueressig \cite{deWS}. They considered
$\cF^{A,B}$ as a general function obeying certain consistency conditions
and defined $\cF^{A, B, C}{}_{ij}$ via \eqref{eq_F2b}. Here we will show
that their consistency conditions are indeed satisfied by the construction
\eqref{eq_F1}.

The first set of consistency conditions are
\begin{align}\label{eq_Fconsistency}
\cF^{A,B,C}{}_{ij} = \cF^{A,C,B}{}_{ij} ~,\qquad
\epsilon^{jk} \cF^{A,B,C}{}_{ij}{}^D{}_{kl} = 0 
\end{align}
and it is very easy to see that these are satisfied by \eqref{eq_F2a}.
The first of them is trivial, while the second follows from
\begin{align}
\cF^{A,B,C}{}_{ij}{}^D{}_{kl} = 
     \frac{1}{8\pi}  \oint_C v^m {\rm d} v_m \,v_i v_j v_k v_l\,
     \frac{\partial^3 \cF^A}{\partial G^{++}_B \partial G^{++}_C \partial G^{++}_D} 
\end{align}
and the property that $\eps^{jk} v_j v_k = 0$.

The second set of conditions, required for the superconformal case,
are\footnote{These were originally written in \cite{deWS} as a single
equation, but here we will consider them separately.}
\begin{align}
\label{eq_Fconsistency2}
\cF^{A, B, C}{}_{ik} G^{k i}_C &= -\cF^{A,B} ~,\\ 
\label{eq_Fconsistency3}
\cF^{A, B, C}{}_{k (i} G_C{}^{k}_{j)} &= 0~.
\end{align}
The first of these follows from the fact that $\cF^A$ 
is a homogeneous function of degree zero and so
\begin{align}
\frac{\partial \cF^A}{\partial G_C^{++}} G_C^{++} = 0 ~.
\end{align}
Taking a  derivative with respect to $G_B^{++}$, we find
\begin{align}\label{eq_temp3}
\frac{\partial^2 \cF^A}{\partial G_B^{++} \partial G_C^{++}} G_C^{++} = -\frac{\partial \cF^A}{\partial G_B^{++}}~.
\end{align}
Then \eqref{eq_Fconsistency2} follows since
\begin{align}
\cF^{A, B, C}{}_{ik} G^{k i}_C =
     \frac{1}{8\pi}  \oint_C v^k {\rm d} v_k \, G_C^{++}
     \frac{\partial^2 \cF^A}{\partial G^{++}_B \partial G^{++}_C}
     = - \cF^{A,B}
\end{align}
using \eqref{eq_temp3}.
To prove \eqref{eq_Fconsistency3} is a little trickier. One may begin
by introducing a fixed isotwistor $u_i$ and writing the condition as
$u^i u^j \cF^{A, B, C}{}_{k i} G_C{}^{k}_{j} = 0$. Then we note that
\begin{align}
u^i u^j \cF^{A, B, C}{}_{k i} G_C{}^{k}_{j}
     &= \frac{1}{8\pi}  \oint_C v^k {\rm d} v_k \,u^i u^j v_i v_k G_C{}^k_j
     \frac{\partial^2 \cF^A}{\partial G^{++}_B \partial G^{++}_C} \eol
     &= \frac{1}{8\pi}  \oint_C v^k {\rm d} v_k \,(v,u)^2 \, G_C^{+ -}
     \frac{\partial^2 \cF^A}{\partial G^{++}_B \partial G^{++}_C}  \eol
     &= \frac{1}{16\pi}  \oint_C v^k {\rm d} v_k \,\ID^{--}
     \left((v,u)^2 \, \frac{\partial \cF^A}{\partial G^{++}_B} \right)
\end{align}
The right-hand side is a total contour derivative and so it must vanish,
which implies \eqref{eq_Fconsistency3}.

\section{Adding $\cO(2n)$ multiplets}

In this section we construct reduced chiral multiplets out of $\cO(2n)$ multiplets.

\subsection{The case of a single $\cO(2n)$ multiplet}
We consider next a more general projective Lagrangian of the form
\begin{align}\label{eq_O2naction}
\cL^{++} = \frac{\cQ^{(2n)}}{(G^{++})^{n-1}} = G^{++} \, \frac{\cQ^{(2n)}}{(G^{++})^{n}} ~.
\end{align}
where $\cQ^{(2n)}$ is a real covariant $\cO(2n)$ multiplet \cite{KLRT-M2} having the functional form 
\begin{align}
\cQ^{(2n)} = \cQ^{i_1 \cdots i_{2n}}
     v_{i_1} \cdots v_{i_{2n}}~, \qquad
     (\cQ^{i_1 \cdots i_{2n}})^* = \cQ_{i_1 \cdots i_{2n}}
\end{align}
and obeying the analyticity constraints
\begin{align}
\cD_\alpha^+ \cQ^{(2n)} = \cD_\dalpha^+ \cQ^{(2n)} = 0 ~.
\end{align}
We note that $\cQ^{(2)} \equiv \cQ^{++}$ is a tensor multiplet.

The reduced chiral superfield which we construct from \eqref{eq_O2naction} is
\begin{align}\label{eq_O2ncontour}
\mathbb W_n = \frac{1}{8\pi} \oint_C v^{i} \rd v_i \, \Big((\BcD^-)^2 + 4 \bar S^{--}\Big)
     \left(\frac{\cQ^{(2n)}}{({G}^{++})^n}\right)~.
\end{align}
Expanding this out gives
\begin{align}
\mathbb W_n &= \frac{1}{8\pi} \oint_C v^{i} \rd v_i \, \Bigg\{
     \frac{(\BcD^-)^2 \cQ^{(2n)}}{({G}^{++})^n}
     - 2n  \frac{\BcD_\dalpha^- \cQ^{(2n)} \BcD^{\dalpha -} {G}^{++}}{({G}^{++})^{n+1}} \eol
     & 
     - n \cQ^{(2n)} \, \frac{(\BcD^{-})^2 {G}^{(2)}}{({G}^{++})^{n+1}}
     + n (n+1) \cQ^{(2n)} \,\frac{\BcD_\dalpha^{-} {G}^{++} \BcD^{\dalpha -} {G}^{++}}{({G}^{++})^{n+2}}
     + 4 \bar S^{--}\, \frac{\cQ^{(2n)}}{({G}^{++})^n}
     \Bigg\}~.
\end{align}
As before, it turns out that all of the explicit $u$-dependence in this expression
can be removed using the analyticity properties. For a general $\cO(2n)$ multiplet,
it is a straightforward exercise to show that
\begin{subequations}
\bea
\BcD_\dalpha^- \cQ^{(2n)} 
     &=&  \frac{2n}{2n+1} \, \bar \eta_\dalpha^{(2n-1)}~,\\
  (\BcD^-)^2 \cQ^{(2n)}
     + 8 n S^{--} \cQ^{(2n)}
     - 4 S^{-+} \ID^{--} \cQ^{(2n)}
     &= &\left(\frac{2n-1}{2n+1}\right) \bar \cH^{(2n-2)} ~,
\eea
\end{subequations}
where
\begin{subequations}
\bea
\bar \eta^\dalpha{}^{(2n-1)} &:=& \BcD^\dalpha_k
     \cQ^{k \,i_1 \cdots i_{2n-1}} v_{i_1} \cdots v_{i_{2n-1}}~, \label{5.7a} \\
\bar {\cH}^{(2n-2)} &:=& \left(\BcD_{jk} + 4 (2n+1) \bar S_{jk} \right)
     \cQ^{j k \,i_1 \cdots i_{2n-2}} v_{i_1} \cdots v_{i_{2n-2}}~.
\label{5.7b}
\eea
\end{subequations}
These are, of course, generalizations of equations in 
subsection \ref{Evaluation of the contour integral}
involving the $\cO(2)$ multiplet ${G}^{++}$.
We will need those other results, too. Applying these relations, we find
\begin{align}\label{eq_O2nExplicit}
\mathbb W_n &= \frac{1}{8\pi} \oint_C v^{i} \rd v_i \, \Bigg\{
     \frac{2n-1}{2n+1}\frac{\bar\cH^{(2n-2)}}{({G}^{++})^n}
     - \frac{8n^2}{3 (2n+1)}  \frac{\bar\eta^{(2n-1)} \bar\chi^+}{({G}^{++})^{n+1}} \eol
     & \qquad
     - \frac{n}{3} \bar M\, \frac{\cQ^{(2n)}}{({G}^{++})^{n+1}}
     + \frac{4 n (n+1)}{9} \,\frac{\cQ^{(2n)} \,\bar\chi^+ \bar\chi^+}{({G}^{++})^{n+2}}
     \Bigg\}~,
\end{align}
where we have again eliminated a total derivative term
\begin{align}
-\frac{1}{8\pi} \oint_C v^{i} \rd v_i \, \ID^{--} \left(4 S^{-+} \frac{\cQ^{(2n)}}{({G}^{++})^n}\right) = 0~.
\end{align}

Now we apply eq. \eqref{eq_contour2} to each of the four terms in \eqref{eq_O2nExplicit}.
The result is a rather unwieldy expression:
\begin{align}
\mathbb W_n =
     -\frac{(2n)!}{2^{2n+1}\, (n!)^2} &\Bigg\{
     \frac{n}{2 (2n+1)} \,
     \frac{\bar\cH^{i_1 \cdots i_{2n-2}}
     {G}_{(i_1 i_2} \cdots {G}_{i_{2n-3} i_{2n-2})}}{{G}^{2n-1}} \eol
     &\quad
     - \frac{2n^2}{3(2n+1)} \,
       \frac{\bar\eta^{i_1 \cdots i_{2n-1}} \bar\chi^{i_{2n}}
     {G}_{(i_1 i_2} \cdots {G}_{i_{2n-1} i_{2n})}}{{G}^{2n+1}} \eol
     &\quad
     - \frac{n}{12} \,
     \frac{\bar M \cQ^{i_1 \cdots i_{2n}}
     {G}_{(i_1 i_2} \cdots {G}_{i_{2n-1} i_{2n})}}{{G}^{2n+1}} \eol
     &\quad
     + \frac{n (2n+1)}{18} \,
     \frac{\cQ^{i_1 \cdots i_{2n}} \bar\chi^{i_{2n+1}} \bar\chi^{i_{2n+2}}
     {G}_{(i_1 i_2} \cdots {G}_{i_{2n+1} i_{2n+2})}}{{G}^{2n+3}} \Bigg\}     ~.
\end{align}
However, as with the improved tensor action, there is a simpler, more compact expression
which is equivalent to this one:
\begin{align}\label{eq_In}
\mathbb W_n
     &= -\frac{(2n)!}{2^{2n+2}\, (n+1)! (n-1)!}\,
     {G} \, (\BcD_{ij} + 4 \bar S_{ij}) \mathcal R^{ij}_n~,
\end{align}
where
\begin{align}\label{eq_O2nR}
\mathcal R_n^{ij} =
     \left(\delta^{ij}_{kl} - \frac{1}{2 {G}^2} {G}^{ij} {G}_{kl} \right)
     \cQ^{kl \,i_1 \cdots i_{2n-2}}
      {G}_{i_1 i_2} \cdots {G}_{i_{2n-3} i_{2n-2}}  {G}^{-2n} ~.
\end{align}
The expression for $\mathbb W_n$ has an overall structure quite similar to
the improved tensor action result \eqref{eq_imptensor}, except the argument
$\mathcal R_n^{ij}$ of the derivative is much more complicated.

Thankfully, many of these complications may be easily understood.
The factor in parentheses in \eqref{eq_O2nR} is simply an
orthogonal projector on the $ij$ indices; it ensures in particular
that if we choose $\cQ^{(2n)} = ({G}^{++})^n$, we get $\mathbb W_n = 0$,
which follows trivially from the contour integration.
Furthermore, a nontrivial check on the combinatoric factors of \eqref{eq_In} can
be made by considering the replacement $\cQ^{(2n)} \rightarrow \cQ^{(2n-2)} {G}^{++}$, under which we
ought to find $\mathbb W_n \rightarrow \mathbb W_{n-1}$. (This is obvious by
considering the original expression \eqref{eq_O2ncontour} for $\mathbb W_n$.) This is a straightforward
combinatoric exercise, the most difficult step of which is to make the replacement
in \eqref{eq_O2nR} of $\cQ^{kl\, i_1 \cdots i_{2n-2}}$ with
\begin{multline}
\frac{{G}^{kl} \cQ^{i_1 \cdots i_{2n-2}}}{2n^2-n}
     + \frac{2n-2}{2n^2-n} \Big({G}^{k (i_{2n-2}} \cQ^{i_1 \cdots i_{2n-3})\, l}
     + {G}^{l (i_{2n-2}} \cQ^{i_1 \cdots i_{2n-3})\, k} \Big)\\
     + \frac{(n-1) (2n-3)}{2n^2-n} {G}^{(i_{2n-3} i_{2n-2}} \cQ^{i_1 \cdots i_{2n-4})\, kl}~.
\end{multline}
The first term vanishes when contracted with the orthogonal projector.
The second and third terms, when contracted with
${G}_{(i_1 i_2} \cdots {G}_{i_{2n-3} i_{2n-2})} {G}^{-2n}$ sum to
\begin{align}
\frac{2 (n^2-1)}{2n^2-n} \cQ^{kl \,i_1 \cdots i_{2n-4}} {G}_{(i_1 i_2} \cdots {G}_{i_{2n-5} i_{2n-4})} {G}^{-2(n-1)}
\end{align}
the numeric prefactor of which is exactly right to convert the
expression for $\mathbb W_n$ to that of $\mathbb W_{n-1}$.

Our $\cO(2n)$ multiplet $\cQ^{(2n)}$ may be an independent dynamical
variable. If $n>1$, it may be chosen instead to be a composite field. For instance, we can choose 
$\cQ^{(2n)}=\cQ_1^{(2m)}\cQ^{(2n-2m)}_2$, with $\cQ_1^{(2m)}$ and $ \cQ^{(2n-2m)}_2$
being $\cO(2m)$ and $\cO(2n-2m)$  multiplets respectively, $m=1,\dots, n-1$.
Another option is to realize $\cQ(2n)$ as a product of $n$ tensor multiplets $H^{++}_A$,
\bea
\cQ^{(2n)} = H^{++}_1 \dots H^{++}_n~.
\eea

In the  case $n=2$ choosing  $\cQ^{(4)} = (H^{++})^2$  leads to the Lagrangian
\begin{align}
\cL^{++} = \frac{(H^{++})^2}{{G}^{++}}
\end{align}
which is a curved-superspace version of that proposed in \cite{BSiegel,deWRV} to describe the 
classical universal hypermultiplet \cite{CFG}.
Using that result, we have
\begin{align}
\frac{1}{8\pi} \oint_C v^{i} \rd v_i \, \Big((\BcD^-)^2 + 4 \bar S^{--}\Big)
     \Big(\frac{H^{++}}{{G}^{++}}\Big)^2
     = - \frac{G}{16} (\BcD_{ij} + 4 \bar S_{ij}) \mathcal R_2^{ij}~,
\end{align}
where
\begin{align}
\mathcal R_2^{ij} = \frac{1}{ {G}^{4}} \left(\delta^{ij}_{kl} - \frac{1}{2 {G}^2} {G}^{ij} {G}_{kl} \right)
     H^{(kl} H^{m n)} {G}_{mn}~.
\end{align}

In the simplest case  case $n=1$,  $\cQ^{(2)} \equiv H^{++}$  we get 
\begin{align}
{\mathbb W}_1= \frac{1}{8\pi} \oint_C v^{i} \rd v_i \, \Big((\BcD^-)^2 + 4 \bar S^{--}\Big)
      \frac{H^{++}}{{G}^{++}}
     = - \frac{G}{16} (\BcD_{ij} + 4 \bar S_{ij}) \mathcal R_1^{ij}~,
\end{align}
where 
\begin{align}
\mathcal R_1^{ij} = \frac{1}{ {G}^{2}} \Big(H^{ij} 
-  G^{ij} \frac{G \cdot H }{G^2} \Big)~, \qquad 
G \cdot H := \hf G^{kl}H_{kl}~.
\end{align}
The reduced chiral scalar ${\mathbb W}_1$ is such that 
\bea
\int \rd^4 x \,{\rm d}^4\q \, \cE \,\J {\mathbb W}_1 ~+~{\rm c.c.} =0~,
\eea
as a consequence of the identity (see eq. (4.61) in \cite{KT-M2})
\bea
\oint_C v^i \rd v_i
\int \rd^4 x \,{\rm d}^4\q \,{\rm d}^4{\bar \q}
\,E\, \frac{{ W}{\bar { W}} }{({ \S}^{++})^2} H^{++}=0~.
\eea

\subsection{Generalization to several $\cO(2n)$ multiplets}
We consider next a more general projective Lagrangian constructed out of
a set of $\cO(2n_A)$ multiplets $\cQ_A^{(2n_A)}$ and at least one 
tensor multiplet $G^{++}$, specifically
\begin{align}\label{eq_genO2naction}
\cL^{++} = G^{++} \cF(\cQ_A^{(2n_A)})~.
\end{align}
The function $\cF$ is required to be a homogeneous function of degree zero in $v^i$;
this implies
\begin{align}
\cF(c^{n_A} \cQ_A^{(2n_A)}) = \cF(\cQ_A^{(2n_A)})~.
\end{align}
This construction is a generalization of that presented in
Section \ref{several_tensor_multiplets}, which involved only
$\cO(2)$ multiplets.\footnote{In Section \ref{several_tensor_multiplets},
an index $A$ was placed on the function $\cF$ and the tensor multiplet
$G^{++}$ in the corresponding construction \eqref{eq_genTaction} to match the
notation in \cite{deWS}. Here we leave such an index off and consider only a single
function $\cF$ for simplicity.}

The reduced chiral superfield which we construct from \eqref{eq_genO2naction} is
\begin{align}
\mathbb W = \frac{1}{8\pi} \oint_C v^{i} \rd v_i \, \Big((\BcD^-)^2 + 4 \bar S^{--}\Big)
     \cF(\cQ_A^{(2n_A)})~.
\end{align}
Expanding this out and applying \eqref{5.7a} and \eqref{5.7b} gives
\begin{align}
\mathbb W &= \frac{2n_B-1}{2n_B+1} \cF^{A}{}_{i_1 \cdots i_{2n_A-2}}
          \bar \cH_A^{i_1 \cdots i_{2n_A-2}}
     \eol & \quad
     + \left(\frac{2n_A}{2n_A+1}\right) \left(\frac{2n_B}{2n_B+1}\right)
          \cF^{AB}{}_{i_1 \cdots i_{2 (n_A+n_B-1)}} \,
          \bar\eta_A^{i_1 \cdots i_{2n_A-1}} \bar \eta_B^{i_{2n_A} \cdots i_{2(n_A + n_B - 1)}}
\end{align}
where
\begin{align}
\cF^{A}{}_{i_1 \cdots i_{2n_A-2}}
     & := \frac{1}{8\pi}  \oint_C v^k {\rm d} v_k \,
     \frac{\partial \cF}{\partial \cQ^{(2n_A)}_A} v_{i_1} \cdots v_{i_{2n_A-2}} ~, \\
\cF^{A B}{}_{i_1 \cdots i_{2 (n_A+n_B-1)}} &:=
     \frac{1}{8\pi}  \oint_C v^k {\rm d} v_k \,
     \frac{\partial^2 \cF}{\partial \cQ^{(2n_A)}_A \partial \cQ^{(2n_B)}_B}
     v_{i_1} \cdots v_{i_{2 (n_A+n_B-1)}}
\end{align}
are both totally symmetric in their isospin indices, 
and $\bar\eta_A$ and  $\bar\cH_A$ are as defined in 
(\ref{5.7a}) and (\ref{5.7b}) respectively, 
with $\cQ$ replaced by $\cQ_A$.
It is worth noting that the second of these expressions can be written
\begin{align}
\cF^{AB}{}_{i_1 \cdots i_{2n_A-2} \,j_1 \cdots j_{2n_B}}
     = \frac{\partial \cF^{A}{}_{i_1 \cdots i_{2n_A-2}}}
               {\partial \cQ_B^{j_1 \cdots j_{2n_B}}}~.
\end{align}

\section{Discussion}
In this paper we have proposed a construction
to generate reduced chiral superfields 
from covariant projective multiplets, including tensor multiplets, $\cO (2n) $ multiplets, etc.
It is given by the relation  (\ref{2.11a}).
In conjunction with the standard construction to derive $\cN=2$ linear multiplets 
from vector ones, eq. (\ref{Sigma2}), we are now able to generate nontrivial higher derivative 
couplings. 
For simplicity,  we illustrate the idea by considering models with vector and tensor multiplets. 

We can start  from a system of tensor multiplets $G^{++}_A$, where $A=1,\dots, n$, and 
introduce a function $\cF_{\rm tensor} (G^{++}_A)$ which is homogeneous  of degree zero, 
\bea
G^{++}_A \frac{\pa  }{\pa G^{++}_A} \cF_{\rm tensor} =0~.
\eea
Then the following superfield
\begin{align}
\mathbb W = \frac{1}{8\pi}  \oint_C v^i {\rm d} v_i
 \Big( ({\bar \cD}^-)^2 +4 \bar{S}^{--}\Big) \cF_{\rm tensor} (G^{++}_A )
\label{6.1}
\end{align}
is reduced chiral, in accordance with the consideration in section 2.

As a next step, we consider two types of reduced chiral superfields 
 $W_I$ and ${\mathbb W}_{\hat I}$.
Here  $W_I$ is the  field strength of a physical vector multiplet, while 
$\mathbb W_{\hat I}$ is a {\it composite} field strength  
of the form (\ref{6.1}).  We  introduce a function 
 $\cF_{\rm chiral} (W_I, {\mathbb W}_{\hat J} )$ which is  homogeneous of  degree one
\bea
\Big( W^I \frac{\pa  }{\pa W^I}  + 
{\mathbb W}_{\hat J}  \frac{\pa  }{\pa {\mathbb W}_{\hat J}} \Big) \cF_{\rm chiral} =\cF_{\rm chiral}  
\eea
Then the superfield
\begin{align}
\mathbb G^{++} = \frac{1}{4} \Big((\CD^{+})^2 + 4 S^{++}\Big) \cF_{\rm chiral}(W_I, {\mathbb W}_{\hat J})
     ~+~{\rm c.c.}
     \label{6.4}
\end{align}
is an $\cO(2)$ multiplet.

Now, the two procedures described can be repeated. We can consider two types of $\cO(2)$ multiplets, 
tensor multiplets $G^{++}_A$ and composite ones ${\mathbb G}^{++}_{\hat B}$ by the rule (\ref{6.4}).
We next pick a function of these multiplets, $\cF_{\cO(2) } (G^{++}_A , {\mathbb G}^{++}_{\hat B})$,
 which is homogeneous  of degree zero, 
\bea
\Big( G^{++}_A \frac{\pa  }{\pa G^{++}_A} + {\mathbb G}^{++}_{\hat B}
 \frac{\pa}{\pa {\mathbb G}^{++}_{\hat B} }\Big)
\cF_{\cO(2)} =0~.
\eea
Using this function, we generate the following reduced chiral superfield
\begin{align}
\mathbb W = \frac{1}{8\pi}  \oint_C v^i {\rm d} v_i
 \Big( ({\bar \cD}^-)^2 +4 \bar{S}^{--}\Big) \cF_{\cO(2)} (G^{++}_A ,{\mathbb G}^{++}_{\hat B})~.
\label{6.6}
\end{align}
Next, we can make use of this reduced scalar to derive new linear multiplets, and so on and so forth.

Each of the two constructions employed adds two spinor derivatives (or one vector derivative).
This differs from a more traditional way (see, e.g., \cite{dWKvZ} and references therein)
to generate higher derivative structures using the chiral projection 
operator $\bar \D$, eq. (\ref{chiral-pr}).  We recall that given a  scalar, isoscalar and U$(1)_R$-neutral 
superfield $U(z)$, which is inert under the super-Weyl transformations, 
its descendant  $\bar \D U$ has the properties
\bea
{\mathbb J} {\bar \D} U = -4 {\bar \D} U~, \qquad {\bar \cD}^\ad_i {\bar \D} U =0~, 
\qquad \d_\s {\bar \D} U = 2 \s {\bar \D} U~,
\eea
where $\d_\s  {\bar \D} U $ denotes the super-Weyl variation of $ {\bar \D} U $.
Given a vector multiplet $W$ such that $W$ is nowhere vanishing, 
we can define the chiral scalar $W^{-2}  {\bar \D} U$ which is neutral under 
the local U$(1)_R$ and  the super-Weyl transformations. 
The latter superfield can be used to construct an antichiral superfield of the form 
${\bar W}^{-2} \D(W^{-2}  {\bar \D} U)$, which is neutral under 
the local U$(1)_R$ and  the super-Weyl transformations, 
 and so on and so forth.\footnote{This 
is a generalization of the construction \cite{BKT}
 of rigid superconformal invariants containing $F^n$, 
with $F$ the electromagnetic field strength.}

Using these chiral operators, one may construct higher derivative actions 
involving chiral Lagrangians. However, it is usually possible
to convert the chiral Lagrangian, which involves an integral over the chiral
subspace, into an integral over the whole superspace by eliminating one of
the chiral projection operators. Schematically, if
$\cL_c = \Phi \bar\Delta U$ for some chiral superfield $\Phi$ and
a well-defined   local and gauge-invariant operator\footnotemark  $U$, then
\begin{align}
\int \rd^4x\,\rd^4\theta\, \cE\, \Phi \bar\Delta U
     = \int \rd^4x\,\rd^4\theta\,\rd^4\bar\theta\, E\, \Phi \, U~.
\end{align}
Thus, higher derivative actions of this type are invariably most naturally
written as integrals over the entire superspace and are not intrinsically
chiral. This has important ramifications for perturbative
calculations, where non-renormalization theorems place strong restrictions
on intrinsic chiral Lagrangians.

\footnotetext{It is important to assume that $U$ is a 
well-defined local and  gauge-invariant operator
and not, for example, a prepotential; else,
any chiral action may be rewritten as an integral over the full superspace
in this way.}

The constructions we are considering are interesting partly because they
are higher derivative terms which \emph{cannot} be written as full superspace
integrals, at least not without introducing prepotentials. As an example,
let us choose the Lagrangian $\cL^{++}$ in (\ref{InvarAc}) as
 $\cL^{++} = {\mathbb G}^{++}  \cF_{\cO(2)} (G^{++}_A ,{\mathbb G}^{++}_{\hat B}) $, 
where ${\mathbb G}^{++} $ is given by eq. (\ref{6.4}). Upon integration by parts we get
\bea
I &=&
\frac{1}{2\pi} \oint_C v^i \rd v_i
\int \rd^4 x \,{\rm d}^4\q \,{\rm d}^4{\bar \q}
\,E\, \frac{{\bar { W}} }{{ \S}^{++}} \, \O
~,
\label{3/4}
\eea
where we have defined the composite Lagrangian
\bea
\O (W_I, {\mathbb W}_{\hat J}, G^{++}_A ,{\mathbb G}^{++}_{\hat B})
 := \cF_{\rm chiral}(W_I, {\mathbb W}_{\hat J})  \cF_{\cO(2)} (G^{++}_A ,{\mathbb G}^{++}_{\hat B})~.
\eea
The specific feature of this Lagrangian is that it obeys the single constraint 
\bea
{\bar \cD}^+_\ad \O =0 ~.
\eea
Although we have written \eqref{3/4} as an integral over the full superspace,
this is really the locally supersymmetric generalization\footnote{This 
way of writing Lagrangians over subspaces in terms
of the full superspace is familiar from $\cN=1$ superspace, where
there are two ways to write
chiral actions, either as an integral over the chiral subspace \cite{SG},
$\int\, \rd^4x\, \rd^2\theta\, \cE\, \cL_c$, or as an integral over the full superspace \cite{WZ,Zumino78},
$ \int\, \rd^4x\, \rd^2\theta\, \rd^2\bar\theta \frac{E}{R} \cL_c$.
Eq. (\ref{3/4})
is analogous to the second of these forms, and so is the projective action (\ref{InvarAc}). 
The locally supersymmetric version of \eqref{eq_sixThetaGlobal}, analogous to the first form,
has not yet been written down within the approach of \cite{KLRT-M1,K-08,KLRT-M2}.}
of the globally
supersymmetric action
\begin{align}\label{eq_sixThetaGlobal}
I_{\rm rigid} =- \frac{1}{8\pi} \oint_C v^i \rd v_i
\int \rd^4 x \,
({\bar D}^-)^2 D^4 
\, \O~, \qquad {\bar D}^+_\ad \O =0 ~,
\end{align}
where the spinor derivatives may be understood as arising from an integration
over six Grassmann coordinates.
For a large class
of such Lagrangians, this action cannot be rewritten as an integral over the
whole superspace of eight Grassmann coordinates without the introduction of
prepotentials. As with $\cN=1$ theories, this has implications for
non-renormalization theorems.

We should point out  that  special holomorphic 
three-derivative contributions
to $\cN=2$ supersymmetric Yang-Mills  effective actions, which are  given as  an integral over 3/4 of superspace, 
have been discussed in the literature \cite{Argyres:2003tg}.

The results of this paper allow us to obtain a simple form for the projective-superspace action \cite{K-08} of 
the minimal formulation for $\cN=2$ Poincar\'e supergravity with vector and tensor 
compensators \cite{deWPV}. Using the techniques developed, 
the gauge-invariant supergravity action can be written as 
\bea
S_{\rm SUGRA}  &=& \frac{1}{ \k^2} \int \rd^4 x \,{\rm d}^4\q \, \cE \, \Big\{
\J {\mathbb W} - \frac{1}{4} W^2 +m \J W \Big\}          +{\rm c.c.}   \non \\
 &=& \frac{1}{ \k^2} \int \rd^4 x \,{\rm d}^4\q \, \cE \, \Big\{
\J {\mathbb W} - \frac{1}{4} W^2 \Big\} +{\rm c.c.}
     ~+~ \frac{m}{ \k^2} \int \rd^4 x \,{\rm d}^4\q \, \rd^4\bar\theta\,
     E \, G^{ij}V_{ij}~,~~~~~~~
\label{6.11} \\
&& \qquad \quad  \mathbb W := -\frac{G}{8} (\BcD_{ij} + 4 \bar S_{ij}) \left(\frac{G^{ij}}{G^2} \right) ~, \non
\eea
where $\k$ is the gravitational constant, $m$ the cosmological constant,  
$\mathbb W$ is given by eq. (\ref{eq_imptensor}), and
$V_{ij}$ is the Mezincescu prepotential \eqref{e.7}.
Within the projective-superspace approach of \cite{KLRT-M1,K-08,KLRT-M2}, 
this action is equivalently given by (\ref{InvarAc})
with the following Lagrangian 
\bea
\k^2 \,\cL^{++}_{\rm SUGRA} =       {G}^{++}  \ln \frac{{ G}^{++}}{{\rm i} \U^+ \breve{\U}{}^+}      
-       \hf {V}\,{\S}^{++} +mV\, G^{++}~,
\label{6.14}
\eea
with $V$ the tropical prepotential for the vector multiplet, and $\U^+$ a weight-one arctic multiplet
(both $\U^+$ and its smile-conjugate $\breve{\U}{}^+$ are pure gauge degrees of freedom).
The first term in the right-hand side of (\ref{6.14}) is (modulo sign) the locally supersymmetric version 
of the projective-superspace Lagrangian for the improved tensor multiplet
constructed in \cite{KLR}. The fact that the vector  and the tensor  multiplets are compensators 
means that their field strengths  $W$ and $G^{ij}$ should possess non-vanishing  expectation values, that is 
$W\neq 0$ and $ G\equiv\sqrt{\frac{1}{2} G^{ij}G_{ij}} \neq 0$.
These conditions are consistent with the equations of motion for the gravitational superfield 
(see \cite{Butter:2010sc} for a recent discussion)
\be
G-W \bar W= 0~.
\label{6.10}
\ee
The equations of motion for the compensators are 
\begin{subequations} 
\bea
\S^{++} -m G^{++} &=&0 ~, \\
\mathbb W + m W &=& 0~.
\label{6.15}
\eea
\end{subequations}

A remarkable feature of the supergravity action (\ref{6.11}) is that its reduction to component
fields can readily be carried out using the technique developed in \cite{GKT-M}.

If the multiplet of conformal supergravity is considered as a curved superspace background, 
the action  (\ref{6.11}) describes (modulo sign) 
a massive vector multiplet or a massive tensor multiplet \cite{K-08}.
The rigid superspace limit of (\ref{6.11}) was introduced 
for the first time  by Lindstr\"om and Ro\v{c}ek 
using $\cN=1$ superfields \cite{LR}.
Their construction was immediately generalized to $\cN=2$ superspace
 \cite{Siegel83} as a simple extension (${\bar D}_{ij} G^{ij} \to \mathbb W$) 
of  the massive $\cN=2$ tensor multiplet model proposed earlier by Howe, Stelle and Townsend
\cite{HST}.  More variant models for massive $\cN=2$ tensor multiplets  
can be found in \cite{K-tensor} . 

The super-Weyl gauge freedom of (\ref{6.11}) can used to fix $G=1$.
The geometric implications of such a gauge fixing have been spelled out in subsection 
\ref{seubsection3-2}. Since $W\neq 0$, the  local U$(1)_R$ freedom 
allows us to impose the  gauge $W-\bar W =0$.

The supergravity theory with Lagrangian (\ref{6.14}) possesses a dual formulation described
solely in terms of a chiral scalar $\J$ and its conjugate $\bar \J$ \cite{K-08}.
Using the techniques developed in the present paper, the dual formulation can be written as  
\bea
S_{\rm SUGRA}  = \frac{1}{ \k^2} \int \rd^4 x \,{\rm d}^4\q \, \cE \, \Big\{
\J {\mathbb W} + \frac{1}{4}\m(\m+\ri e)  \J^2 \Big\} +{\rm c.c.} ~, 
\label{6.17}
\eea
where  $m^2 =\m^2 + e^2$,  with $\m \neq 0$.
Here $\J$ and its conjugate $\bar\J$ are the only conformal compensators. 
The action is super-Weyl invariant, with $\J$ transforming as 
\be
\d_\s \J=\s \J~,
\ee
in spite of the presence of the mass term.
Unlike 
common wisdom, we see that $\cN=2$ Poincar\'e supergravity can be realized without a compensating 
vector multiplet.\footnote{The vector multiplet has been eaten up by the tensor multiplet 
which is now massive. The vector compensator acts as a St\"uckelberg field to give mass to the tensor multiplet.
This is an example of the phenomenon observed originally in \cite{LM} and studied in detail 
in \cite{Dall'Agata:2003yr,DSV,DF,LS,Theis:2004pa,K-tensor,K-08,Gauntlett:2009zw}.}
The complex mass parameter in (\ref{6.17}) can be interpreted 
to have both electric and magnetic contributions
which are associated with the two possible 
mass terms $B \wedge {}^* B$ 
and $B \wedge B$ for the component two-form $B$ (see, e.g, \cite{DF} for a pedagogical discussion).

The action (\ref{6.17}) leads to the following equation of motion for $\J$ 
\bea
\mathbb W + \hf  \m(\m+\ri e)  \J =0~.
\label{6.19}
\eea
Provided eq. (\ref{6.19}) holds, the equation of motion for the gravitational superfield is 
\bea
G- \m^2 \J \bar \J =0~,
\eea
compare with (\ref{6.15}). 
The latter implies that $\J$ is nowhere vanishing on-shell. We therefore are allowed to impose 
the super-Weyl gauge $G=1$ and 
fix the local U$(1)_R$ symmetry as $\J - \bar \J=0$.
\\

\noindent
{\bf Acknowledgements:}\\
Discussions with Gabriele Tartaglino-Mazzucchelli are gratefully acknowledged.
This work  is supported in part by the Australian Research Council 
and by a UWA Research Development Award.

\appendix

\section{Geometry of conformal supergravity}
\label{Appendix A}
\noindent 
This section is taken essentially verbatim from \cite{KLRT-M2}. 

We give a summary of the superspace geometry for $\N=2$
conformal supergravity  which was originally  introduced  in \cite{Howe}, as a generalization of \cite{Grimm}, 
and later elaborated 
in \cite{KLRT-M2}.
A curved four-dimensional  $\cN=2$ superspace  $\cM^{4|8}$ is parametrized by
local  coordinates  $z^{{M}}=(x^{m},\q^{\mu}_\imath,{\bar \q}_{\dot{\mu}}^\imath)$,
where $m=0,1,\cdots,3$, $\mu=1,2$, $\dot{\mu}=1,2$ and  $\imath= \1,\2$.
The Grassmann variables $\q^{\mu}_\imath $ and $\teb_{\dot{\mu}}^\imath$
are related to each other by complex conjugation: 
$\overline{\q^{\mu}_\imath}=\teb^{\dot{\mu}\imath}$. 
The structure group is ${\rm SL}(2,{\mathbb C})\times {\rm SU}(2)_R \times {\rm U}(1)_R$,
with $M_{ab}=-M_{ba}$, $J_{ij}=J_{ji}$ and $\mathbb J$ be the corresponding 
Lorentz, ${\rm SU}(2)_R$ and ${\rm U}(1)_R$ generators.
The covariant derivatives 
$\cD_{{A}} =(\cD_{{a}}, \cD_{{\a}}^i,\cDB^\ad_i) 
\equiv (\cD_{{a}}, \cD_{ \underline{\a} }, \cDB^{\underline{\ad}})$ 
have the form 
\bea
\cD_{A}&=&E_{A}+
\hf \,\O_{A}{}^{bc}\,M_{bc}+
\Phi_{A}^{~\,kl}\,J_{kl}
+ \ri \,\Phi_{A}\,{\mathbb J} \non\\
&=&E_{A}~+~\O_{A}{}^{\b\g}\,M_{\b\g}
+{\O}_{A}{}^{\bd\gd}\,\bar{M}_{\bd\gd}
+\Phi^{~\,kl}_{A}\,J_{kl}
+\ri \, \Phi_{A}\,{\mathbb J}~.
\label{CovDev}
\eea
Here $E_{{A}}= E_{{A}}{}^{{M}} \pa_{{M}}$ is the supervielbein, 
with $\pa_{{M}}= \pa/ \pa z^{{M}}$,
$\O_{{A}}{}^{bc}$ is the Lorentz connection, 
 $\Phi_{{A}}{}^{kl}$ and $\Phi_{{A}}$ are  
 the ${\rm SU}(2)_R$ and ${\rm U}(1)_R$ connections, respectively.

The Lorentz generators with vector indices ($M_{ab}$) and spinor indices
($M_{\a\b}=M_{\b\a}$ and ${\bar M}_{\ad\bd}={\bar M}_{\bd\ad}$) are related to each other 
by the standard rule:
$$
M_{ab}=(\s_{ab})^{\a\b}M_{\a\b}-(\tilde{\s}_{ab})^{\ad\bd}\bar{M}_{\ad\bd}~,~~~
M_{\a\b}=\hf(\s^{ab})_{\a\b}M_{ab}~,~~~
\bar{M}_{\ad\bd}=-\hf(\tilde{\s}^{ab})_{\ad\bd}M_{ab}~.
$$ 
The generators of the structure group
act on the spinor covariant derivatives as follows:\footnote{The 
(anti)symmetrization of $n$ indices 
is defined to include a factor of $(n!)^{-1}$.}
\bea
{[}M_{\a\b},\cD_{\g}^i{]}
&=&\ve_{\g(\a}\cD^i_{\b)}~,\qquad
{[}\bar{M}_{\ad\bd},\cDB_{\gd}^i{]}=\ve_{\gd(\ad}\cDB^i_{\bd)}~, \non \\
{[}J_{kl},\cD_{\a}^i{]}
&=&-\d^i_{(k}\cD_{\a l)}~,
\qquad
{[}J_{kl},\cDB^{\ad}_i{]}
=-\ve_{i(k}\cDB^\ad_{l)}~, \non \\
{[}{\mathbb J},\cD_{\a}^i{]} &=&\cD_{\a}^i~,\qquad  \qquad ~~\,
{[}{\mathbb J},\cDB^{\ad}_i{]}~
=\,-\cDB^{\ad}_i~,
\label{generators}
\eea
Our notation and conventions correspond to \cite{BK}.

The  covariant derivatives obey the algebra
{\allowdisplaybreaks
\begin{subequations}\label{eq_algebra}
\bea
\{\cD_\a^i,\cD_\b^j\}&=&
4S^{ij}M_{\a\b}
+2\ve^{ij}\ve_{\a\b}Y^{\g\d}M_{\g\d}
+2\ve^{ij}\ve_{\a\b}\bar{W}^{\gd\dd}{\bar M}_{\gd\dd}
\non\\
&&
+2 \ve_{\a\b}\ve^{ij}S^{kl}J_{kl}
+4 Y_{\a\b}J^{ij}~,
\label{a-c1}\\
\{\cD_\a^i,\cDB^\bd_j\}&=&
-2\ri\d^i_j(\s^c)_\a{}^\bd\cD_c
+4 \big( \d^{i}_{j}G^{\d\bd}  +\ri G^{\d\bd}{}^{i}{}_{j} \big) M_{\a\d}
+4\big( \d^{i}_{j}G_{\a\gd}
+\ri G_{\a\gd}{}^{i}{}_{j}\big) {\bar M}^{\gd\bd} ~~~~~~
\non\\
&&
+8 G_\a{}^\bd J^{i}{}_{j}
-4\ri\d^i_jG_{\a}{}^\bd{}^{kl}J_{kl}
-2\big( \d^i_jG_{\a}{}^{\bd}
+\ri G_{\a}{}^{\bd}{}^i{}_j\big) {\mathbb J} ~, 
\label{a-c2}\\
{[}\cD_a,\cD_\b^j{]}&=& -\ri 
(\ts_a)^{\ad\g}\Big(
\d^j_kG_{\b\ad}
+ \ri G_{\b\ad}{}^{j}{}_k\Big)
\cD_\g^k
\non\\
&&
+{\frac\ri 2}\Big(({\s}_a)_{\b\gd}S^{jk}
-\ve^{jk}({\s}_a)_\b{}^{\dd}\bar{W}_{\dd\gd}
-\ve^{jk}({\s}_a)^{\a}{}_\gd Y_{\a\b}\Big)\cDB^\gd_k
\non\\
&&
+\hf R_a{}_\b^j{}^{cd}M_{{c}{d}}
+R_a{}_\b^j{}^{kl}J_{kl}
+\ri R_a{}_\b^j\,{\mathbb J} ~.
\label{vector-spinor}
\eea
\end{subequations}
}
Here  the dimension-1 components of the torsion
obey the symmetry properties  
\bea
S^{ij}=S^{ji}~, \qquad Y_{\a\b}=Y_{\b\a}~, 
\qquad W_{\a\b}=W_{\b\a}~, \qquad G_{\a\ad}{}^{ij}=G_{\a\ad}{}^{ji}
\eea
and the reality conditions
\bea
\overline{S^{ij}} =  \bar{S}_{ij}~,\quad
\overline{W_{\a\b}} = \bar{W}_{\ad\bd}~,\quad
\overline{Y_{\a\b}} = \bar{Y}_{\ad\bd}~,\quad
\overline{G_{\b\ad}} = G_{\a\bd}~,\quad
\overline{G_{\b\ad}{}^{ij}} = ~G_{\a\bd}{}_{ij}.
\eea
The ${\rm U}(1)_R$ charges of the complex fields are:
\bea
{\mathbb J} \,S^{ij}=2S^{ij}~,\qquad
{\mathbb J}  \,Y_{\a\b}=2Y_{\a\b}~, \qquad
{\mathbb J} \, W_{\a\b}=-2W_{\a\b}~, \qquad
{\mathbb J}  \, W=-2W~.
\eea
The dimension-3/2 components of the curvature appearing in (\ref{vector-spinor}) 
have the following explicit form:
\begin{subequations}
\bea
R_a{}_\b^j{}_{cd}&=&
-\ri(\s_d)_{\b}{}^{\dd} T_{ac}{}_\dd^j
+\ri(\s_a)_{\b}{}^{\dd} T_{cd}{}_\dd^j
-\ri(\s_c)_{\b}{}^{\dd} T_{da}{}_\dd^j
~,
\label{3/2curvature-1}
\\
R_{\a\ad}{}_{\b}^j{}^{kl}
&=&
-{\ri}\ve^{j(k}\cDB_\ad^{l)}Y_{\a\b}
-{\ri}\ve_{\a\b}\ve^{j(k}\cDB^{\dd l)}\bar{W}_{\ad\dd}
-{\frac\ri 3}\ve_{\a\b}\ve^{j(k}\cDB_{\ad q}S^{l)q}
\non\\
&&
+{\frac43}\ve^{j(k}\cD_{(\a q}G_{\b)\ad}{}^{l)q}
+{\frac23}\ve_{\a\b}\ve^{j(k}\cD^\d_{q}G_{\d\ad}{}^{l)q}
~,
\\
R_{\a\ad}{}_{\b}^j&=&
-\cD_{\b}^jG_{\a\ad}
+{\frac\ri 3}\cD_{(\a k}G_{\b)\ad}{}^{jk}
+{\frac\ri 2}\ve_{\a\b}\cD^{\g}_kG_{\g\ad}{}^{jk}~.
\eea
\end{subequations}
The right-hand side of  (\ref{3/2curvature-1}) involves the dimension-3/2 components 
of the torsion which are expressed in terms of the dimension-1 tensors as follows: 
\begin{subequations}
\bea
&&T_{ab}{}_\gd^k\equiv(\s_{ab})^{\a\b}\cT_{\a\b}{}_{\gd}^{k}
-(\ts_{ab})^{\ad\bd}\cT_{\ad\bd}{}_{\gd}^{k}~,~~~
\\
&&\cT_{\a\b}{}_{\gd}^{k}
=
-{\frac14}\cDB_{\gd}^{k}Y_{\a\b}
+{\frac\ri 3}\cD_{(\a}^lG_{\b)\gd}{}^{k}{}_l
~,
\\
&&\cT_{\ad\bd}{}_{\gd}^{ k}
=
-{\frac14}\cDB_{\gd}^k\bar{W}_{\ad\bd}
-{\frac16}\ve_{\gd(\ad}\cDB_{\bd)l}S^{kl}
-{\frac\ri 3}\ve_{\gd(\ad}\cD^{\d}_qG_{\d\bd)}{}^{kq}
~.
\eea
\end{subequations}

The dimension-3/2 Bianchi identities are:
\begin{subequations}\label{eq_dim3.5bianchi}
\bea
\cD_{\a}^{(i}S^{jk)}&=&0~, \qquad 
\cDB_{\ad}^{(i}S^{jk)} = \ri\cD^{\b (i}G_{\b\ad}{}^{jk)}~,
\label{BI-3/2-1}
 \\
\cD_\a^i\bar{W}_{\bd\gd}&=&0~,\\
\cD_{(\a}^{i}Y_{\b\g)}&=&0~, \qquad 
\cD_{\a}^{i}S_{ij}+\cD^{\b}_{j}Y_{\b\a}=0~, \\
\cD_{(\a}^{(i}G_{\b)\bd}{}^{jk)}&=&0~, \\
\cD_\a^iG_{\b\bd}&=&
- \frac{1}{ 4}\cDB_\bd^iY_{\a\b}
+ \frac{1}{ 12}\ve_{\a\b}\cDB_{\bd j}S^{ij}
- \frac{1}{ 4}\ve_{\a\b}\cDB^{\gd i}\bar{W}_{\gd\bd}
- \frac{\ri }{ 3}\ve_{\a\b}\cD^{\g}_j G_{\g \bd}{}^{ij}~.
\eea
\end{subequations}

\section{Chiral projection operator} 
Actions in $\cN=2$ supergravity may be constructed from integrals
over the full superspace
\begin{align}
\int \rd^4 x \,{\rm d}^4\q \,{\rm d}^4{\bar \q}
\,E\, \cL
\end{align}
or integrals over a chiral subspace
\begin{align}
\int \rd^4 x \,{\rm d}^4\q \,
\,\cE\, \cL_c ~, \qquad {\bar \cD}^\ad_i \cL_c =0
\end{align}
with $\cE$ the chiral density.
Just as in $\cN=1$ superspace, actions of the former type
may be rewritten as the latter using a covariant chiral projection
operator $\bar \D$ \cite{Muller},
\begin{align}
\int \rd^4 x \,{\rm d}^4\q \,{\rm d}^4{\bar \q}
\,E\, \cL = \int \rd^4 x \,{\rm d}^4\q \,
\,\cE\, \bar\D \cL~.
\label{A.12}
\end{align}
The  covariant chiral projection operator is defined as 
\bea
\bar{\D}
&=&\frac{1}{96} \Big((\cDB^{ij}+16\bar{S}^{ij})\cDB_{ij}
-(\cDB^{\ad\bd}-16\bar{Y}^{\ad\bd})\cDB_{\ad\bd} \Big)
\non\\
&=&\frac{1}{96} \Big(\cDB_{ij}(\cDB^{ij}+16\bar{S}^{ij})
-\cDB_{\ad\bd}(\cDB^{\ad\bd}-16\bar{Y}^{\ad\bd}) \Big)~.
\label{chiral-pr}
\eea
Its fundamental property  is that $\bar{\D} U$ is covariantly chiral,
for any scalar, isoscalar and U$(1)_R$-neutral superfield $U(z)$,
\be
{\bar \cD}^{\ad}_i \bar{\D} U =0~.
\ee
A detailed derivation of the relation (\ref{A.12}) can be found in \cite{KT-M2}.

It follows from the explicit structure of the  chiral projection operator that
\bea
\int \rd^4 x \,{\rm d}^4\q \,{\rm d}^4{\bar \q}
\,E\, \F =0~,
\eea
for any covariantly chiral scalar $\F$ of zero U$(1)_R$ charge, 
${\bar \cD}^\ad_i \F = {\mathbb J} \F =0$.

\section{Isotwistors and projective superspace}
In this paper, our isotwistor notation and conventions differ slightly from those adopted in 
\cite{KLRT-M1,K-08,KLRT-M2}, but agree with those used in \cite{Kuzenko:2010bd}.

Associated with any completely symmetric ${\rm SU}(2)$ tensor 
$V^{i_1 \cdots i_n} = V^{( i_1 \cdots i_n )}$ 
is a superfield
$V^{(n)}$ obeying
\begin{align}
V^{(n)} = V^{i_1 \cdots i_n} v_{i_1} \cdots v_{i_n} ~,
\end{align}
with the $(n)$ superscript referring to the degree of homogeneity
in the isotwistor parameter $v_i$. We often have need to introduce
an additional isotwistor $u_i$, which is linearly independent of $v_i$
\begin{align}
(v,u) = v^k u_k = \eps^{kj} v_j u_k \neq 0
\end{align}
in terms of which we may define new isotwistors $V^{(n-m, m)}$
\begin{align}
V^{(n-m,m)} = V^{i_1 \cdots i_n} v_{i_1} \cdots v_{i_{n-m}} 
\frac{u_{i_{n-m+1}}}{(v,u)} \cdots \frac{u_{i_n}}{(v,u)}
\end{align}
with degree $n-2m$. For the cases of $n=1$ and $n=2$, we will use a more
compact notation involving $+$ and $-$:
\begin{gather}
V^{+} = V^i v_i, \qquad
V^{-} = V^i \frac{u_i}{(v,u)} \\
V^{++} = V^{i j} v_i v_j, \quad
V^{+-} = V^{i j} v_i \frac{u_j}{(v,u)}, \quad
V^{--} = V^{i j} \frac{u_i u_j}{(v,u)^2}
\end{gather}

Because the covariant derivatives of $\cN=2$ superspace have isospin
indices we may identify
\begin{gather}
\cD_\alpha^+ = v_i \cD_\alpha^i ~,\qquad
\BcD_\dalpha^+ = v_i \BcD_\dalpha^i \\
\cD_\alpha^- = \frac{u_i}{(v,u)} \cD_\alpha^i ~,\qquad
\BcD_\dalpha^- = \frac{u_i}{(v,u)} \BcD_\dalpha^i
\end{gather}
It will be useful to introduce derivative operations on the isotwistors
themselves,
\begin{align}
\ID^{--} = \frac{1}{(v,u)} \, u^i \frac{\partial}{\partial v^i}~, \qquad
\ID^{++} = (v,u) \, v_i \frac{\partial}{\partial u_i}
\end{align}
which have the properties that
\begin{gather}
\ID^{++} V^{(n-m,m)} = m V^{(n-m+1,m-1)}, \qquad
\ID^{--} V^{(n-m,m)} = (n-m) V^{(n-m-1,m+1)} ~.
\end{gather}

For those knowledgable of harmonic superspace \cite{GIOS}, the above definitions 
will seem familiar. They can be derived from corresponding objects in harmonic
superspace by the formal replacements
\begin{gather}
u_i^+ \rightarrow v_i ~,\qquad
u_i^- \rightarrow \frac{u_i}{(v,u)} ~.
\end{gather}
The difference from the harmonic superspace definitions is that the isotwistor $v^i$ is not normalized, and
 $u_i$ is not related to $v^i$ by complex conjugation.

\section{Contour integrals in $\mathbb CP^1$}\label{residue}
In this paper, we have need to evaluate contour integrals of the general form
\begin{align}
\mathcal C_n = \oint_C v^{i} \rd v_i \, \frac{\Omega^{(2n-2)}}{(G^{++})^n}~.
\end{align}
In this expression $\Omega^{(2 n - 2)}$ is some homogeneous function of
degree $2n-2$ in the variable isotwistor $v^i$, degree zero in the fixed
isotwistor $u_i$, and obeying the analyticity constraints
\begin{align}
\cD_\alpha^+ \Omega^{(2n-2)} = \cD_\dalpha^+ \Omega^{(2n-2)} = 0 ~.
\end{align}
We further assume that the contour encloses a region where the only
singularities are those arising from the $(G^{++})^n$ factor in the
denominator.

Under these assumptions, the contour may be evaluated using a trick familiar from
twistor theory. We write $G^{ij} = \ri \,\omega^{(i} \bar \omega^{j)}$ in terms of isotwistors
$\omega^i$ and $\bar\omega^j$. Then
\begin{align}
G^2 = \frac{1}{2} G^{ij} G_{ij} = \frac{1}{4} (\omega^j \bar \omega_j)^2
\Longrightarrow G = \frac{1}{2} \omega^j \bar\omega_j
\end{align}
since the reality of $G^{ij}$ implies $\bar\omega_j = (\omega^j)^*$.
So long as $G$ is nonzero, $\omega^j$ and $\bar\omega^j$ are linearly independent
isotwistors, in terms of which the contour integral may be written
\begin{align}
\mathcal C_n = \oint_C \frac{v^{i} \rd v_i}{(\ri \, \omega^{+} \bar\omega^{+})^n} \, \Omega^{(2n-2)} ~,
\end{align}
where
\begin{align}
\omega^+ = \omega^i v_i, \quad
\bar\omega^+ = \bar\omega^i v_i ~.
\end{align}
The pole in the contour appears when either $\omega^{+}$ or $\bar\omega^{+}$ vanishes --
that is, when either $v^{i} \propto \omega^i$ or $\bar\omega^i$.
Because $G \neq 0$, these poles are distinct and we can consider their residues separately.

Without loss of generality, we will restrict to the case where the contour
encircles $\omega^i$. The $n$th order pole may be converted to a first order
pole using the relation
\begin{align}\label{eq_pole}
(\ID^{--})^{j} \frac{1}{\omega^{+}} = \frac{(-1)^j j! (\omega^{-})^j}{(\omega^{+})^{j+1}}
\end{align}
where
\begin{align}
\ID^{--} &\equiv \frac{u^i}{(v,u)} \frac{\partial}{\partial v^i} \\
\omega^{-} &\equiv \frac{\omega^i u_i}{(v,u)}
\end{align}
and applying integration by parts. To do this, we first rewrite the contour integral
using \eqref{eq_pole}:
\begin{align}
\mathcal C_n
     &= \frac{(-1)^{n-1}}{(n-1)!} \oint_C v^{i} \rd v_i \,\frac{1}{(\omega^{-})^{n-1}}
     \left((\ID^{--})^{n-1} \frac{1}{\omega^{+}} \right)
     \frac{\Omega^{(2n-2)}}{(\ri \, \bar\omega^{+})^n} ~.
\end{align}
Next we would like to flip each of the $\ID^{--}$ operators off the pole.
For this step to be valid, we must check that the total derivative
terms actually do vanish. Each of them has the form
\begin{align}
\oint_C v^{i} \rd v_i \, \ID^{--} \cF
\end{align}
where $\cF = \cF(v, u)$ is a function of degree
zero in the isotwistors $v^i$ and $u_i$ separately.
Because of the homogeneity property, we may trade $v^i$ and $u_i$
for projective coordinates $\zeta$ and $\xi$ where
\begin{align}
v^i = v^{\ul 1} (1, \zeta), \qquad
u_i = u_{\ul 1} (1, \xi)
\end{align}
with $\cF$ depending only on $\zeta$ and $\xi$, and
the contour integral rewritten as
\begin{align}
\oint_C v^{i} \rd v_i \, \ID^{--} \cF = \oint_C \rd\zeta \,\partial_\zeta \cF
     = \oint_C \rd t \, \dot\zeta \, \partial_\zeta \cF~,
\end{align}
where in the second equality we have parametrized the contour with a real variable $t$.
Because $u_i$ is fixed, $\xi$ is independent of $t$ and the integrand
is a total derivative in $t$, so the contour vanishes.

Noting that the operator $\ID^{--}$ annihilates $\omega^{-}$, the term
generated by integrating by parts is\footnote{There is a subtlety in this procedure.
$\ID^{--}$ annihilates $1 / \omega^{-}$ only if $u_i$
is chosen to be linearly independent of $\omega_i$.
This is analogous to functions of a single complex variable where
$\bar\partial \,(z - z_0)^{-1} = 0$ only for $z \neq z_0$.}
\begin{align}
\mathcal C_n
     &= \frac{1}{(n-1)!} \oint_C \frac{v^{i} \rd v_i}{\omega^{+}} \frac{1}{(\omega^-)^{n-1}}
     (\ID^{--})^{n-1} \left(\frac{\Omega^{(2n-2)}}{(\ri\, \bar\omega^{+})^n}\right)~.
\end{align}
In evaluating the $\ID^{--}$'s on their argument, we would like to eliminate
all terms coming from $\ID^{--}$ hitting the $\bar\omega^{+}$ factors in the denominator.
This is possible if we choose $u_i = \bar\omega_i$:
\begin{align}\label{eq_contour}
\mathcal C_n
     &= \frac{1}{(n-1)!} \oint_C \frac{v^{i} \rd v_i}{\omega^{+}}
     \left(\frac{(\ID^{--})^{n-1}\Omega^{(2n-2)}}{(\omega^{-})^{n-1} (\ri\, \bar\omega^{+})^n}\right)
     \Big\vert_{u_i = \bar\omega_i}~.
\end{align}

Having reduced our expression to a first order pole, we now apply the
residue theorem. Given a contour integral
\begin{align}
\oint_C \frac{v^{i} \rd v_i}{\omega^j v_j}\, \cF^{-}(v)~,
\end{align}
with $\cF^{-}$ a homogeneous function of $v^i$ of degree $-1$
which is nonsingular at $v^i \propto \omega^i$, we may rewrite it
in terms of the inhomogeneous coordinate $\zeta = v^{\ul 2} / v^{\ul 1}$.
If we exploit the freedom to choose $v^{\ul 1} = w^{\ul 1}$, then
\begin{align}
v^i = w^{\ul 1} (1, \zeta)~.
\end{align}
This leads to
\begin{align}
\oint_C \frac{d\zeta}{-\omega^{\ul 2} / \omega^{\ul 1} + \zeta}
     \, \cF^{-}(w^{\ul 1}(1,\zeta))
     = 2\pi \ri \, 
     \cF^{-} (\omega^i) ~.
\end{align}
We have assumed the contour to be evaluated in a counterclockwise fashion,
but in principle the opposite sign may also arise.

Applying this result to \eqref{eq_contour} gives 
\begin{align*}
\mathcal C_n
     &= \frac{2\pi \ri}{(n-1)!}
     \left(\frac{(\ID^{--})^{n-1}\Omega^{(2n-2)}}{(\omega^{-})^{n-1} (\ri\, \bar\omega^{+})^n}\right)
     \Big\vert_{u_i = \bar\omega_i, v^i = \omega^i} ~.
\end{align*}
The terms in the denominator may be simplified by noting that
\begin{gather*}
\bar\omega^{+} = -2 G ,\qquad
\omega^{-} = 1
\end{gather*}
giving
\begin{align}
\mathcal C_n &\equiv \oint_C v^{i} \rd v_i \, \frac{\Omega^{(2n-2)}}{(G^{++})^n}
     = \frac{2\pi \, \ri^{n+1}}{(n-1)! (2 G)^n}
     \left((\ID^{--})^{n-1}\Omega^{(2n-2)}\right)\Big\vert_{\bar v_i = \bar\omega_i, v^i = \omega^i} ~.
\end{align}

${}$For the cases of interest to us, $\Omega^{(2j)}$ is of the form
$\Omega^{i_1 \cdots i_{2j}} v_{i_1} \cdots v_{i_{2j}}$. Then
$\ID^{--}$ may be evaluated explicitly to give
\begin{align}
(\ID^{--})^{j}\Omega^{(2j)} &= \frac{1}{(v,u)^j} \, \frac{(2j)!}{j!} \Omega^{i_1 \cdots i_{2j}}
     v_{(i_1} \cdots v_{i_{j}} u_{i_{j+1}} \cdots u_{i_{2j})}  \non \\
     &= \frac{(-\ri)^j (2j)!}{2^j j!} \, \Omega^{i_1 \cdots i_{2j}}
     G_{(i_1 i_2} \cdots G_{i_{2j-1} i_{2j})} G^{-j}
\end{align}
where we have taken $v^i = \omega^i$ and $u_i = \bar\omega_i$.
This gives our main result
\begin{align}\label{eq_contour2}
\mathcal C_n
     &= -\frac{2\pi}{2^{2n-1}}
     \, \frac{(2n-2)!}{(n-1)! (n-1)!}\,
     \Omega^{i_1 \cdots i_{2n-2}} G_{(i_1 i_2} \cdots G_{i_{2n-3} i_{2n-2})} G^{-(2n-1)}~.
\end{align}

\section{Prepotential formulations for vector multiplet}\label{Wprepotential}
Within the  projective-superspace approach of \cite{KLRT-M1,K-08,KLRT-M2}, the 
constraints on the vector multiplet field strength
$W$ can be solved  in terms of a real weight-zero tropical prepotential $V(v^i)$
\cite{KT-M} as in eq. (\ref{1.4}).
Here we use this construction to introduce a curved-superspace analogue 
of  Mezincescu's prepotential \cite{Mezincescu}.

First of all, let us show how Mezincescu's prepotential for the vector multiplet can be introduced 
within standard superspace.
For this a simple generalization of  the rigid supersymmetric analysis in \cite{HST} can be used.
One begins with the first-order  action
\begin{align}
S &= \frac{1}{4} \int \rd^4 x \,{\rm d}^4\q \,\cE\, \cW \cW
+     \frac{1}{4} \int \rd^4 x \,{\rm d}^4\bar\q \,\bar\cE\, \bar \cW \bar \cW \eol
     &\qquad
     - \frac{\rm i}{8} \int \rd^4 x \,{\rm d}^4\q \,{\rm d}^4{\bar \q}
\,E\, \Big(\cW (\CD^{ij} + 4 S^{ij}) V_{ij} - \bar \cW (\BCD^{ij} + 4 \bar S^{ij}) V_{ij}\Big)~,
\label{F1}
\end{align}
where $\cW$ is a covariantly chiral superfield, and $V^{ij}=V^{ji} $ is an {\it unconstrained} real
SU(2) triplet  acting as a Lagrange multiplier. 
Varying (\ref{F1}) with respect to $V_{ij}$ gives $\cW =W$, where $W$ obeys the Bianchi identity 
(\ref{1.1}). As a result, the second term in (\ref{F1}) drops out and we end up with the Maxwell action
\bea
S = \frac{1}{2} \int \rd^4 x \,{\rm d}^4\q \,\cE\, W W ~.
\eea
On the other hand,
because the action (\ref{F1}) is quadratic in $\cW$,
we may easily integrate $\cW$ out using its equation of motion
\begin{align}\label{eq_WMezincescu}
\cW =  {\rm i} W_{\rm D}~,  \qquad W_{\rm D}:= \frac{1}{4}\bar\Delta (\CD^{ij} + 4 S^{ij}) V_{ij} ~.
\end{align}
This leads to the dual action
\begin{align}
S = \frac{1}{2} \int \rd^4 x \,{\rm d}^4\q \,\cE\, W_{\rm D} W_{\rm D} ~.
\end{align}
The dual field strength  $W_{\rm D}$ must be both reduced chiral and given by \eqref{eq_WMezincescu}.

We now show how  to construct the Mezincescu prepotential $V_{ij}$
within projective superspace. One begins with the expression for
$W$ in terms of a weight-zero tropical prepotential  $V(v^i)$, eq. (\ref{1.4}). 
The analyticity conditions on $V$ may be solved 
in terms of  an unconstrained isotwistor superfield $\cU^{(-4)}$
(see \cite{KLRT-M1} for the definition of isotwistor superfields), which is real under the smile-conjugation,
as follows
\begin{align}
V &= \frac{1}{16} \Big(({\bar \cD}^+)^2 +4 \bar{S}^{++}\Big) \Big(({\cD}^+)^2 + 4 S^{++}\Big)  \cU^{(-4)} \eol
     &= \frac{1}{16} \Big(({\cD}^+)^2 + 4 S^{++}\Big) \Big(({\bar \cD}^+)^2 +4 \bar{S}^{++}\Big) \cU^{(-4)}~.
\end{align}
Using this construction, one may write $W$, using the results of \cite{KT-M2},  as
\begin{align}
W &= \frac{1}{128\pi} \oint_C v^i {\rm d}v_i
     \Big( ({\bar \cD}^-)^2 +4 \bar{S}^{--}\Big)
     \Big(({\bar \cD}^+)^2 +4 \bar{S}^{++}\Big) \Big(({\cD}^+)^2 + 4 S^{++}\Big)  \cU^{(-4)} \eol
     &= \frac{\bar\Delta}{8\pi} \oint_C v^i {\rm d}v_i \Big(({\cD}^+)^2 + 4 S^{++}\Big)  \cU^{(-4)}
\end{align}
where $\bar\Delta$ is the chiral projection operator \eqref{chiral-pr}.
This may subsequently be rewritten
\begin{align}\label{e.7}
W &= \frac{\bar\Delta}{8\pi} \, \Big({\cD}^{ij} + 4 S^{ij}\Big)  \,\oint_C v^k {\rm d}v_k \, v_i v_j\, \cU^{(-4)} \eol
     &= \frac{1}{4}\bar\Delta \Big({\cD}^{ij} + 4 S^{ij}\Big) V_{ij}~,
\end{align}
where we have defined the Mezincescu prepotential
\begin{align}
V_{ij} = \frac{1}{2\pi} \oint_C v^k {\rm d}v_k \, v_i v_j\, \cU^{(-4)} ~.
\end{align}
This construction for $W$ is manifestly chiral due to the appearance
of the projection operator. To prove the Bianchi identity, one
may start with \eqref{1.4} and replace the dummy integration variable
$v^i \to \hat{v}{}^i$ and rewrite the expression for $W$ in
the form
\bea
{W}  &=& \frac{1}{8\pi}  \oint_C 
\frac{ \hat{v}{}^i {\rm d}\hat{v}_i}{(\hat{v},u)^2}\,
u_i u_j  \Big( {\bar \cD}^{ij} +4 \bar{S}^{ij}\Big) V (\hat{v}) ~. 
\eea  
Since this expression does not depend on the constant isospinor $u_i$, we can choose 
it as $u_i \propto v_i$, and then the last expression turns into
\bea
{W}  &=& 
 \Big( ({\bar \cD}^+)^2 +4 \bar{S}^{++}\Big)
\oint_C 
\frac{ \hat{v}{}^i {\rm d}\hat{v}_i}{8\p (\hat{v},v)^2}\, V (\hat{v}) ~. 
\eea  
We now can check the fulfilment of the Bianchi identity:
 \bea
&& \Big( ({ \cD}^+)^2 +4 {S}^{++}\Big){W} -  \Big( ({\bar \cD}^+)^2 +4 \bar{S}^{++}\Big)
\bar{W} \non \\
&=& \Big[ ({ \cD}^+)^2 +4 {S}^{++} ,   ({\bar \cD}^+)^2 +4 \bar{S}^{++}\Big] 
\oint_C 
\frac{ \hat{v}{}^i {\rm d}\hat{v}_i}{8\p (\hat{v},v)^2}\, V (\hat{v})  \\
& \equiv & \Big[ ({ \cD}^+)^2 +4 {S}^{++} ,   ({\bar \cD}^+)^2 +4 \bar{S}^{++}\Big]  \U~.
\non 
\eea 
Since $\U$ is a Lorentz scalar and is neutral under U$(1)_R$, we find
\begin{align}
& \Big[ (\cD^{++})^2 + 4 S^{++}, (\BcD^{++})^2 + 4 \bar S^{++} \Big] \U 
     = 8 (\cD^{\alpha +} \bar S^{++} ) \cD_\alpha^+ \U 
       - 8( \bar\cD_{\dalpha}^+ S^{++} ) \bar\cD^\dalpha{}^+ \U \eol
     & 
     + 8\ri (\cD^{\alpha +} G_{\alpha \dalpha}^{++} )\BcD^{\dalpha +} \U
     + 8\ri (\BcD^{\dalpha +} G_{\alpha \dalpha}^{++}) \cD^{\alpha +} \U 
     + 4 \Big((\cD^{+})^2 \bar S^{++} - (\BcD^+)^2 S^{++}\Big) \U
\end{align}
vanishes when we apply some of the constraints \eqref{eq_dim3.5bianchi},
\begin{align}
\cD_\alpha^+ \bar S^{++} = i \BcD^\dalpha G_{\alpha \dalpha}^{++} ~, \qquad
\BcD_\dalpha^+ \bar S^{++} = i \cD^\alpha G_{\alpha \dalpha}^{++} ~.
\end{align}
and make use of the algebra of covariant derivatives \eqref{eq_algebra} to show
\begin{align}
(\CD^+)^2 \bar S^{++} - (\BCD^+)^2 S^{++} =
     i \{\CD^{\alpha +}, \BCD^{\dalpha +}\} G_{\alpha \dalpha}^{++} = 0 ~.
\end{align}

\footnotesize{

}

\end{document}